\documentclass[twocolumn]{aastex62}
\usepackage{graphicx}
\usepackage{subfigure}
\usepackage{bm}
\usepackage{amsmath}
\usepackage{fixltx2e}
\usepackage{xcolor}
\usepackage{amssymb}
\usepackage{url}

\usepackage[normalem]{ulem}   
\usepackage{color}

\makeatletter
\newcommand{\rmnum}[1]{\romannumeral #1}
\newcommand{\Rmnum}[1]{\expandafter\@slowromancap\romannumeral #1@}
\makeatother

\begin{document}

\title{Magnetohydrodynamics with GAMER}

\correspondingauthor{Tzihong Chiueh}
\email{chiuehth@phys.ntu.edu.tw}

\author{Ui-Han Zhang}
\affiliation{Department of Physics, National Taiwan University, 10617, Taipei, Taiwan}
\nocollaboration

\author{Hsi-Yu Schive}
\affiliation{National Center for Supercomputing Applications, Urbana, IL, 61801, USA}
\nocollaboration

\author{Tzihong Chiueh}
\affiliation{Department of Physics, National Taiwan University, 10617, Taipei, Taiwan}
\affiliation{Institute of Astrophysics, National Taiwan University. 10617, Taipei, Taiwan}
\affiliation{Center for Theoretical Physics, National Taiwan University, 10617, Taipei, Taiwan}
\nocollaboration

\begin{abstract}
\label{abstract}
GAMER, a parallel Graphic-processing-unit-accelerated Adaptive-MEsh-Refinement hydrodynamic code, has been extended to support magnetohydrodynamics (MHD) with both the corner-transport-upwind (CTU) and MUSCL-Hancock schemes and the constraint transport (CT) technique. The divergent preserving operator for adaptive mesh refinement (AMR) has been applied to reinforce the divergence-free constraint on the magnetic field.  GAMER-MHD has fully exploited the concurrent executions between the GPU MHD solver and other CPU computation pertinent to AMR. We perform various standard tests to demonstrate that GAMER-MHD is both second-order accurate and robust, producing results as accurate as those given by high-resolution uniform-grid runs. We also explore a new 3D MHD test, where the magnetic field assumes the Arnold-Beltrami-Childress (ABC) configuration, temporarily becomes turbulent with  current sheets and finally settles to a lowest-energy equilibrium state. This 3D problem is adopted for the performance test of GAMER-MHD. The single-GPU performance reaches $1.2\times 10^8$ and $5.5\times 10^7$ cell-updates/sec for the single- and double-precision calculations, respectively, on Tesla P100. We also demonstrate a parallel efficiency of $\sim 70\%$ for both weak and strong scaling using $1,024$ XK nodes on the Blue Waters supercomputers.
\end{abstract}


\section{Introduction}
\label{sec:introduction}

Magnetohydrodynamics (MHD) plays crucial roles in many astrophysical settings, such as dynamo of the primordial magnetic field \citep*{Kulsrud1992, Kulsrud1997, Naoz2013, Schober2013}, active galactic nuclei disks \citep*{Balbus1991, Balbus2003, Kazanas2012, Fukumura2015, Ryan2017}, jets \citep*{Blandford1982, Pelletier1992, Ferreira1997, Pudritz2012, Stepanovs2014}, star forming clouds \citep*{ChiuehChou1994, Li1996, Chiueh1998, BalsaraWard-ThompsonCrutcher2001, Shu2004, McKee2007}, and solar interior and atmosphere \citep*{ChiuehZweibel1987, ChiuehZweibel1989, CattaneoChiuehHughes1990_1, CattaneoChiuehHughes1990_2, Chiueh2000}. Many of these problems are intrinsically three dimensional tackled numerically with 3D simulations. However, high-precision 3D simulations are costly.  For example, a reasonable research-grade simulation with $4096^3$ resolution conventionally must run on supercomputers with more than $100$ nodes.

In the past decades, several central-processing-unit (CPU) based codes have been developed for astrophysical applications, for example, ATHENA \citep*{Stone2008}, AstroBEAR \citep*{Cunningham2009}, CHARM \citep*{Miniati2007, Miniati2011} and PLUTO \citep*{Mignone2012}. More recently, a new development has opted in an alternative direction for scientific computing, the graphic-process-unit (GPU) computing. Examples include ENZO \citep*{Bryan2014}, RAMSES \citep*{Kestener2014}, CHOLLA \citep*{Schneider2015}, CLAMR \citep*{Tumblin2015},  GPUPEGAS \citep*{Kulikov2014}, SMAUG \citep*{Grifﬁths2015}, FARGO3D \citep*{Benitez-Llambay2016} and FLASH \citep*{Lukat2016}. One of the earliest achievements in high-performance computing making good use of GPU's capability is GAMER \citep*{bSc1, bSc2}, which is a 3D hydrodynamic (HD) code supporting both a relaxing total variation diminishing (RTVD) scheme \citep*{Jin1995, Trac2003}, a dimensional split and Riemann-solver-free scheme, and several dimensional unsplit and Riemann-solver-based schemes \citep*{bSc2}, for example, the corner-transport-upwind (CTU) scheme \citep*{Colella1990} and the MUSCL-Hancock scheme described by \citet{Falle1991} and \citet{vanLeer2006} (hereafter referred to as the VL scheme). It takes advantage of GPU acceleration for number crunching and differs from other pure GPU codes (e.g., \citet{Schive2008}) in that it also makes good use of the CPU.  Operations other than number crunching in GAMER, such as mesh refinements, data preparation, data transfer among different nodes, Hilbert curve construction, etc., are conducted in the much more versatile, but slower, CPU (this management is different from Daino, a framework proposed by \citet{Wahib2015} and \citet{Wahib2016}, in which the mesh refinement is also conducted in GPU).  As GPU and CPU computing tasks are performed in parallel, concurrency of the two is a strong requirement for good performance, and GAMER can often achieve more than $90\%$ concurrency in different tests \citep{bSc2}. Hence GAMER allows one to efficiently conduct research-grade simulations with a dozen of computing nodes, and has been used to investigate collapse of molecular cloud cores \citep*{Zhang2015} and jets from active galactic nuclei \citep*{Molnar2017}. Not only that, GAMER can also scale up to run in a supercomputer with thousands of nodes \citep*{Schive2017}.

This work extends GAMER to MHD, where we closely follow the GAMER data structure and parallelization scheme. We implement the GPU MHD scheme using the CTU and VL schemes, both of which are HD schemes extended to support MHD in the ATHENA code \citep*{Stone2008}. We also adopt the constraint transport (CT) technique \citep*{Evans1988} to solve the induction equation, a technique that preserves the divergence-free property of the magnetic field and has been implemented in most of the aforementioned codes \citep*{Fromang2006, Stone2008, Cunningham2009, Bryan2014, Miniati2011, Mignone2012, Benitez-Llambay2016}.

One important feature of GAMER is adaptive mesh refinement (AMR), which allows one to increase the resolution in regions of interest dynamically. This technique has been implemented in various astrophysical codes (e.g., \citet{Fryxell2000}, \citet{Fromang2006}, \citet{Cunningham2009}, \citet{Bryan2014}, \citet{Miniati2011}, \citet{Mignone2012} and \citet{Tumblin2015}).  The AMR structure in GAMER is based on constructing a hierarchy of grid patches with an octree data structure similar to the FLASH code \citep*{Fryxell2000}. To satisfy the divergence-free constraint on the magnetic field with AMR, we follow the scheme proposed by \citet{Balsara2001} when interpolating the coarse-grid magnetic field.

By taking advantage of AMR, we can apply sufficient resolution to examine standard test problems.  Here we choose linear waves, two shock-tube problems (\citet{Torrilhon2003} and \citet{Ryu1995}), \citet{Orszag1979} vortex, and the blast wave \citep*{Londrillo2000} as our test problems.  With sufficiently high resolution, we discover several interesting features not reported before.  These new features should be understood via appropriate underlying physical mechanisms and are warranted for separate detailed studies.  In this paper, we will only show robust numerical results and leave the comprehensive analyses in future works.

Inspired by the intriguing flow structure of the Arnold-Beltrami-Childress (ABC) flow \citep*{Arnold1965, Childress1970}, we investigate an ABC magnetic-field configuration where the 3D incompressible flow is replaced by the 3D divergent-free magnetic field in a uniform plasma.  The magnetic field configuration is in force-free equilibrium and the ABC field pattern is controlled by an integer number which determines how many periods appearing in a given box.  The two-period configuration is chosen for investigation, as the one-period configuration is shown in this work to be linearly stable.   The simulation result reveals that the strong current density takes place in sheets, and the two-period configuration relaxes to the one-period force-free equilibrium, in agreement with the Taylor's conjecture \citep*{Taylor1986}.  This 3D problem is also used for the performance test.

This paper is organized as follows. The MHD equations are introduced in Sec. (\ref{sec:MHD Equations}). In Sec. (\ref{sec:Numerical Algorithm}) we present the MHD scheme and the AMR structure. In Sec. (\ref{sec: GAMER MHD Structure}) we describe the hybrid MPI/OpenMP/GPUs implementation. In Sec. (\ref{sec: numerical results}) we show numerical results including  both accuracy and performance tests. Conclusions are made in Sec. (\ref{sec: conclusion}).  Finally, the proof of the stability of the one period ABC-flow magnetic field configuration is presented in Appendix.

\section{MHD Equations}
\label{sec:MHD Equations}

The equations for MHD can be written in conservative form as

\begin{equation}
\label{equ: mass conservation}
{{\partial \rho}\over{\partial t}}+ \nabla \cdot(\rho\bm{V}) = 0,
\end{equation}

\begin{equation}
\label{equ: momentum conservation}
{{\partial \rho\bm{V}}\over{\partial t}}+ \nabla \cdot(\rho\bm{V}\bm{V}-\bm{B}\bm{B} + P^{*}\bm{\Rmnum{1}}) = \bm{0},
\end{equation}

\begin{equation}
\label{equ: energy conservation}
{{\partial e}\over{\partial t}}+ \nabla \cdot[(e+P^{*})\bm{V}-\bm{B}(\bm{B}\cdot\bm{V})] = 0,
\end{equation}

\begin{equation}
\label{equ: induction equation}
{{\partial \bm{B}}\over{\partial t}} + \nabla \times\bm{E} = \bm{0}.
\end{equation}
where $\bm{\Rmnum{1}}$ is the identity tensor, $P^{*}=P+\bm{B}\cdot\bm{B}/2$ with the gas pressure $P$, $\bm{E}=-\bm{V}\times\bm{B}$ is the electric field and $e$ is the total energy density

\begin{equation}
\label{equ: definition of energy}
e = {{P}\over{\gamma-1}} + {{1}\over{2}}\rho\bm{V}\cdot\bm{V}+{{\bm{B}\cdot\bm{B}}\over{2}}.
\end{equation}
Other symbols have their usual meanings. These equations are written in units such that the magnetic permeability $\mu=1$ and the speed of light $c=1$. In addition to these equations, the magnetic field must obey the divergence-free constraint, i.e. $\nabla \cdot \bm{B}=\bm{0}$.

The above equations can be rewritten as ${\partial}_t\bm{U} + {\partial}_x\bm{F}_x + {\partial}_y\bm{F}_y + {\partial}_z\bm{F}_z = \bm{0}$ in Cartesian coordinates. Here $\bm{U}$ is the vector of conserved densities with the following form
\begin{equation}
\label{equ: definition of conserved variables}
\bm{U}=
\begin{bmatrix}
\rho \\
 M_x \\
 M_y \\
 M_z \\
 e   \\
 B_x \\
 B_y \\
 B_z
\end{bmatrix}
,
\end{equation}
where $\bm{M} = (M_x, M_y, M_z) \equiv \rho \bm{V}$ is the momentum density. $\bm{F}_x$,  $\bm{F}_y$ and $\bm{F}_z$ are the fluxes in the $x$-, $y$- and $z$-directions, respectively,

\begin{equation}
\label{equ: definition of F_x}
\bm{F}_x=
\begin{bmatrix}
\rho V_x \\
\rho V_x^2 + P + {1\over2}(\bm{B}\cdot\bm{B}) - B_x^2 \\
\rho V_xV_y - B_xB_y\\
\rho V_xV_z - B_xB_z \\
(e+P^{*})V_x - (\bm{B}\cdot\bm{V})B_x   \\
0 \\
B_yV_x-B_xV_y \\
B_zV_x-B_xV_z
\end{bmatrix}
,
\end{equation}
\begin{equation}
\label{equ: definition of F_y}
\bm{F}_y=
\begin{bmatrix}
\rho V_y \\
\rho V_yV_x - B_yB_x\\
\rho V_y^2 + P + {1\over2}(\bm{B}\cdot\bm{B}) - B_y^2 \\
\rho V_yV_z - B_yB_z \\
(e+P^{*})V_y - (\bm{B}\cdot\bm{V})B_y   \\
B_xV_y-B_yV_x \\
0 \\
B_zV_y-B_yV_z
\end{bmatrix}
,
\end{equation}
\begin{equation}
\label{equ: definition of F_z}
\bm{F}_z=
\begin{bmatrix}
\rho V_z \\
\rho V_zV_x - B_zB_x\\
\rho V_zV_y - B_zB_y \\
\rho V_z^2 + P + {1\over2}(\bm{B}\cdot\bm{B}) - B_z^2 \\
(e+P^{*})V_z - (\bm{B}\cdot\bm{V})B_z   \\
B_xV_z-B_zV_x \\
B_yV_z-B_zV_y \\
0
\end{bmatrix}
.
\end{equation}

\section{Numerical Algorithm}
\label{sec:Numerical Algorithm}

In this section, we describe the MHD algorithm and the AMR scheme implemented in GAMER-MHD.

\subsection{Mathematical Formulation}
\label{subsec: Mathematical Formulation}

\begin{figure}
\includegraphics[scale=0.68]{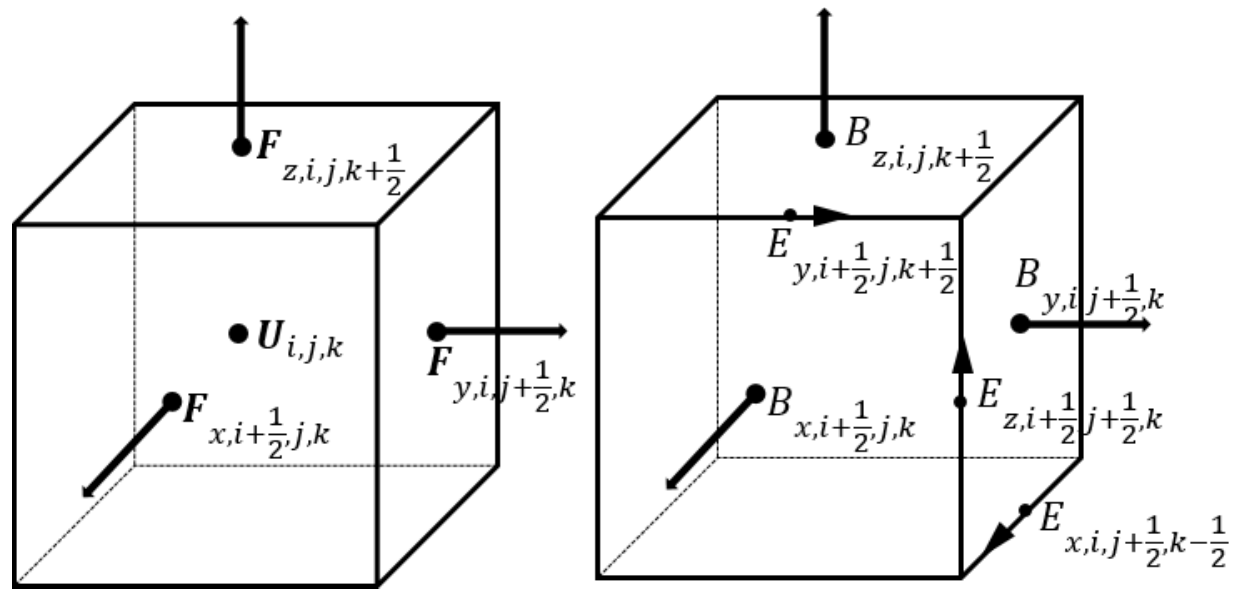}
\caption{Schematic diagram for the locations of conserved variables, fluxes, and the magnetic and electric fields. The flux $\bm{F}$ and the magnetic field $\bm{B}$ are defined on the faces of the cubic cell, the electric field $\bm{E}$ on the cell edges, and the conserved variable $\bm{U}$ at the cell center.}
\label{fig: schematic_for_variables_location}
\end{figure}

The conservative MHD equations given in Sec. (\ref{sec:MHD Equations}) is usually solved by the finite-volume (FV) method, in which conserved variables, such as mass density, momentum density and energy density, are treated as the volume-averaged quantities in a given cubic cell and evolved by fluxes defined on the cell interfaces.  In GAMER-MHD, we adopt both CTU and VL schemes with the CT method. On the other hand, to preserve the divergence-free constraint on the magnetic field, the CT technique proposes that the magnetic field should be treated as the area-averaged quantity on the cell interface and it is evolved by the electric fields defined on the edges of the cubic cell. Furthermore, due to the different treatments of the magnetic field and conserved variables, the ghost zone width should be increased by one cell compared with the pure hydrodynamical case. We shall give a brief review of the FV method, the CT technique and CTU and VL schemes below.

We first introduce notations. Consider a finite domain of size $(L_x, L_y, L_z)$ in each direction, respectively. The continuous spatial coordinate $(x, y, z)$ are discretized into $(N_x, N_y, N_z)$ cells within the domain in each direction. In GAMER, the size of cell in each direction on a given AMR level is uniform throughout the domain, i.e., $\delta x \equiv L_x/N_x, \delta y \equiv L_y/N_y, \delta z \equiv L_z/N_z$ with $\delta x=\delta y=\delta z$. For convenience, the cell center coordinate is denoted as $(x_i, y_j, z_k)$, where $x_i = (i+1/2)\delta x, y_j = (j+1/2)\delta y, z_k = (k+1/2)\delta z$ with $i,j,k \in \mathbb{N}\cup\{ 0 \}$, and $\mathbb{N}$ is the set of all positive integer numbers.

Similarly, time is discretized into a series of time slices. For convenience, we use the superscript to denote the discrete time slice and define $\delta t^n \equiv t^{n+1}-t^{n}$  with $n \in \mathbb{N}\cup\{ 0 \}$. Hereafter, we drop the superscript on $\delta t$.

The FV method is based on the integral form of conservation laws. By integrating the conservation forms of MHD equations over the volume of one cell and a discrete interval of time  $\delta t$, one arrives, after using the Stokes theorem, at

\begin{equation}
\label{equ: finite volume integral form}
\left.\begin{aligned}
\bm{U}^{n+1}_{i,j,k} = \bm{U}^{n}_{i,j,k} & - {{\delta t}\over{\delta x}} \Big ( \bm{F}^{n+{1\over2}}_{x,i+{1\over2},j,k}-\bm{F}^{n+{1\over2}}_{x,i-{1\over2},j,k} \Big ) \\
& - {{\delta t}\over{\delta y}} \Big ( \bm{F}^{n+{1\over2}}_{y,i,j+{1\over2},k}-\bm{F}^{n+{1\over2}}_{y,i,j-{1\over2},k} \Big ) \\
& - {{\delta t}\over{\delta z}} \Big ( \bm{F}^{n+{1\over2}}_{z,i,j,k+{1\over2}}-\bm{F}^{n+{1\over2}}_{z,i,j,k-{1\over2}} \Big ),
\end{aligned}
\right.
\end{equation}
where
\begin{equation}
\label{equ: definition of cell center variables}
\bm{U}^{n}_{i,j,k} \equiv \int^{z_{k+{1\over2}}}_{z_{k-{1\over2}}} \int^{y_{j+{1\over2}}}_{y_{j-{1\over2}}} \int^{x_{i+{1\over2}}}_{x_{i-{1\over2}}} {{\bm{U}(x,y,z,t^n)}\over{\delta x \delta y \delta z}}dxdydz,
\end{equation}
an array of volume-averaged conserved variables, and
\begin{equation}
\left.\begin{aligned}
\label{equ: definition of area center flux}
& \bm{F}^{n+{1\over2}}_{x,i-{1\over2},j,k} \equiv \int^{t^{n+1}}_{t^n} \int^{z_{k+{1\over2}}}_{z_{k-{1\over2}}} \int^{y_{j+{1\over2}}}_{y_{j-{1\over2}}} {{\bm{F}_x \Big ( x_{i-{1\over2}},y,z,t \Big )}\over{\delta t \delta y \delta z}}dydzdt, \\
& \bm{F}^{n+{1\over2}}_{y,i,j-{1\over2},k} \equiv \int^{t^{n+1}}_{t^n} \int^{z_{k+{1\over2}}}_{z_{k-{1\over2}}} \int^{x_{i+{1\over2}}}_{x_{i-{1\over2}}} {{\bm{F}_y \Big ( x,y_{j-{1\over2}},z,t \Big )}\over{\delta t \delta x \delta z}}dxdzdt, \\
& \bm{F}^{n+{1\over2}}_{z,i,j,k-{1\over2}} \equiv \int^{t^{n+1}}_{t^n} \int^{y_{j+{1\over2}}}_{y_{j-{1\over2}}} \int^{x_{i+{1\over2}}}_{x_{i-{1\over2}}} {{\bm{F}_z \Big (x,y,z_{k-{1\over2}},t \Big )}\over{\delta t \delta x \delta y}}dxdydt,
\end{aligned}
\right.
\end{equation}
an array of the time- and area-averaged fluxes.

Here, the half-integer subscripts represent the boundary of the cell and the half-integer superscripts on fluxes are evaluated at the half time step $t^{n+{1\over2}}$.  In this way, the volume-averaged conserved variables can be approximately regarded as variables defined at the cell center and the fluxes are evaluated at the cell boundary face for updating the conserved variables.

The CT technique uses the induction equation (Eq. (\ref{equ: induction equation})) to evolve the magnetic field. Integrating the induction equation over the three orthogonal faces of the cell at $(i-1/2,j,k)$, $(i,j-1/2,k)$ and $(i,j,k-1/2)$ and applying the Stoke's theorem give
\begin{equation}
\label{equ: induction equation integral form}
\left.\begin{aligned}
& B^{n+1}_{x,i-{1\over2},j,k} = B^{n}_{x,i-{1\over2},j,k} & -{{\delta t}\over{\delta y}} \Big ( E^{n+{1\over2}}_{z, i-{1\over2},j+{1\over2},k}-E^{n+{1\over2}}_{z, i-{1\over2},j-{1\over2},k} \Big ) \\
& & +{{\delta t}\over{\delta z}} \Big ( E^{n+{1\over2}}_{y, i-{1\over2},j,k+{1\over2}}-E^{n+{1\over2}}_{y, i-{1\over2},j,k-{1\over2}} \Big ), \\
& B^{n+1}_{y,i,j-{1\over2},k} = B^{n}_{y,i,j-{1\over2},k} & - {{\delta t}\over{\delta z}} \Big ( E^{n+{1\over2}}_{x,i,j-{1\over2},k+{1\over2}}-E^{n+{1\over2}}_{x,i,j-{1\over2},k-{1\over2}} \Big ) \\
& & +{{\delta t}\over{\delta x}} \Big ( E^{n+{1\over2}}_{z, i+{1\over2},j-{1\over2},k}-E^{n+{1\over2}}_{z, i-{1\over2},j-{1\over2},k} \Big ), \\
& B^{n+1}_{z,i,j,k-{1\over2}} = B^{n}_{z,i,j,k-{1\over2}} & - {{\delta t}\over{\delta x}} \Big ( E^{n+{1\over2}}_{y,i+{1\over2},j,k-{1\over2}}-E^{n+{1\over2}}_{y,i-{1\over2},j,k-{1\over2}} \Big ) \\
& & +{{\delta t}\over{\delta y}} \Big ( E^{n+{1\over2}}_{x,i,j+{1\over2},k-{1\over2}}-E^{n+{1\over2}}_{x,i,j-{1\over2},k-{1\over2}} \Big ),
\end{aligned}
\right.
\end{equation}
for which
\begin{equation}
\label{equ: definition of face magnetic field}
\left.\begin{aligned}
& B^{n}_{x,i-{1\over2},j,k} \equiv \int^{z_{k+{1\over2}}}_{z_{k-{1\over2}}} \int^{y_{j+{1\over2}}}_{y_{j-{1\over2}}} {{B_x \Big ( x_{i-{1\over2}},y,z,t^n \Big )}\over{\delta y \delta z}}dydz, \\
& B^{n}_{y,i,j-{1\over2},k} \equiv \int^{x_{i+{1\over2}}}_{x_{i-{1\over2}}} \int^{z_{k+{1\over2}}}_{z_{k-{1\over2}}} {{B_y \Big ( x,y_{j-{1\over2}},z,t^n \Big )}\over{\delta x \delta z}}dxdz, \\
& B^{n}_{z,i,j,k-{1\over2}} \equiv \int^{x_{i+{1\over2}}}_{x_{i-{1\over2}}} \int^{y_{j+{1\over2}}}_{y_{j-{1\over2}}} {{B_z \Big ( x,y,z_{k-{1\over2}},t^n \Big )}\over{\delta x \delta y}}dxdy,
\end{aligned}
\right.
\end{equation}
are area-averaged magnetic field components, and
\begin{equation}
\label{equ: the definition of line electric field}
\left.\begin{aligned}
& E^{n+{1\over2}}_{x,i,j-{1\over2},k-{1\over2}} \equiv \int^{t^{n+1}}_{t^n} \int^{x_{i+{1\over2}}}_{x_{i-{1\over2}}} {{E_x \Big ( x,y_{j-{1\over2}},z_{k-{1\over2}},t \Big )}\over{\delta t \delta x}}dxdt, \\
& E^{n+{1\over2}}_{y,i-{1\over2},j,k-{1\over2}} \equiv \int^{t^{n+1}}_{t^n} \int^{y_{j+{1\over2}}}_{y_{j-{1\over2}}} {{E_y \Big ( x_{i-{1\over2}},y,z_{k-{1\over2}},t \Big )}\over{\delta t \delta y}}dydt, \\
& E^{n+{1\over2}}_{z,i-{1\over2},j-{1\over2},k} \equiv \int^{t^{n+1}}_{t^n} \int^{z_{k+{1\over2}}}_{z_{k-{1\over2}}} {{E_z \Big ( x_{i-{1\over2}},y_{j-{1\over2}},z,t \Big )}\over{\delta t \delta z}}dzdt,
\end{aligned}
\right.
\end{equation}
are electric field components evaluated on cell edges.

Note that as we define the divergent operator in the discrete space to be
\begin{equation}
\label{equ: definition of the divergent operator}
\left.\begin{aligned}
(\nabla \cdot \bm{B})^{n}_{i,j,k} \equiv  & { {B^n_{x,i+{1\over2},j,k} - B^n_{x,i-{1\over2},j,k}}\over {\delta x}} + \\
& { {B^n_{y,i,j+{1\over2},k} - B^n_{y,i,j-{1\over2},k}}\over {\delta y}} + \\
& { {B^n_{z,i,j,k+{1\over2}} - B^n_{z,i,j,k-{1\over2}}}\over {\delta z}},
\end{aligned}
\right.
\end{equation}
it is straightforward to show that $(\nabla \cdot \bm{B})^{n+1}_{i,j,k} = (\nabla \cdot \bm{B})^{n}_{i,j,k}$ by Eq. (\ref{equ: induction equation integral form}). Accordingly, the divergence-free condition of the magnetic field is guaranteed when initially chosen so. Figure (\ref{fig: schematic_for_variables_location}) shows the positions of all averaged variables described above on a cell.

\subsection{MHD Scheme}
\label{subsec: MHD Scheme}

In this section, we summarize the CTU and VL schemes for updating the solution by one time step $\delta t$, which is determined by the Courant-Friedrichs-Lewy stability condition \citep*{Courant1928}. For more details of these two schemes, readers are referred to \citet{Stone2008} and \citet{Stone2009}.

The following procedure is for the CTU scheme:

1. Evaluate the left and right interface values for all cells in all three spatial directions by the 1D data reconstruction. GAMER-MHD supports both the piece-wise linear (PLM) and piece-wise parabolic (PPM) interpolations.

2. Evaluate the fluxes across all cell interfaces by solving the Riemann problem. Here the electric field is defined at the cell interface.  GAMER-MHD supports three Riemann solvers: HLLE, HLLD and Roe solvers.

3. Evaluate the electric field on the cell edge from the electric field at the cell interface obtained by step 2 together with the magnetic field at the cell interface and the velocity at the cell center.

4. Update the area-averaged magnetic field by CT and all volume-averaged conserved variables by the conservative integration for $\delta t/2$.

5. Correct the cell interface value obtained in step 1 by computing the transverse flux gradients.

6. Solve the Riemann problem with the corrected data to obtain the new fluxes across all cell interfaces.

7. Evaluate the electric field on the cell edge again with the electric field obtained by step 6 and the half-step magnetic field and velocity from step 4.

8. Update all conserved variables and the magnetic field by $\delta t$.

On the other hand, the VL scheme can be summarized as follows:

1. Calculate the first-order flux across all cell interfaces by solving the Riemann problem with the longitudinal component of the magnetic field equal to the face-centered value at each interface.

2. Evaluate the electric field on the cell edge. This step is the same as the step 3 in the CTU scheme.

3. Update the area-averaged magnetic field and all volume-averaged conserved variables for $\delta t/2$, which is the same as the step 4 in the CTU scheme.

4. Evaluate the left and right interface values for all cells in all three spatial directions by the 1D data reconstruction at half time step.

5. Solve the Riemann problem with the left and right interface values to obtain the new fluxes across all cell interfaces.

6. Evaluate the electric field on the cell edge again. This step is the same as the step 7 in the CTU scheme.

7. Update all conserved variables and the magnetic field by $\delta t$.

Although both CTU and VL schemes are second-order accurate in space and time, the CTU scheme is generally less diffusive than the VL scheme due to the transverse flux gradient correction (step 5 in the CTU scheme). Hence, we will adopt the CTU scheme for all numerical accuracy tests in Sec. (\ref{subsec: accuracy test}).

\subsection{Adaptive Mesh Refinement}
\label{subsec:Adaptive Mesh Refinement}

\begin{figure}
\includegraphics[scale=0.49]{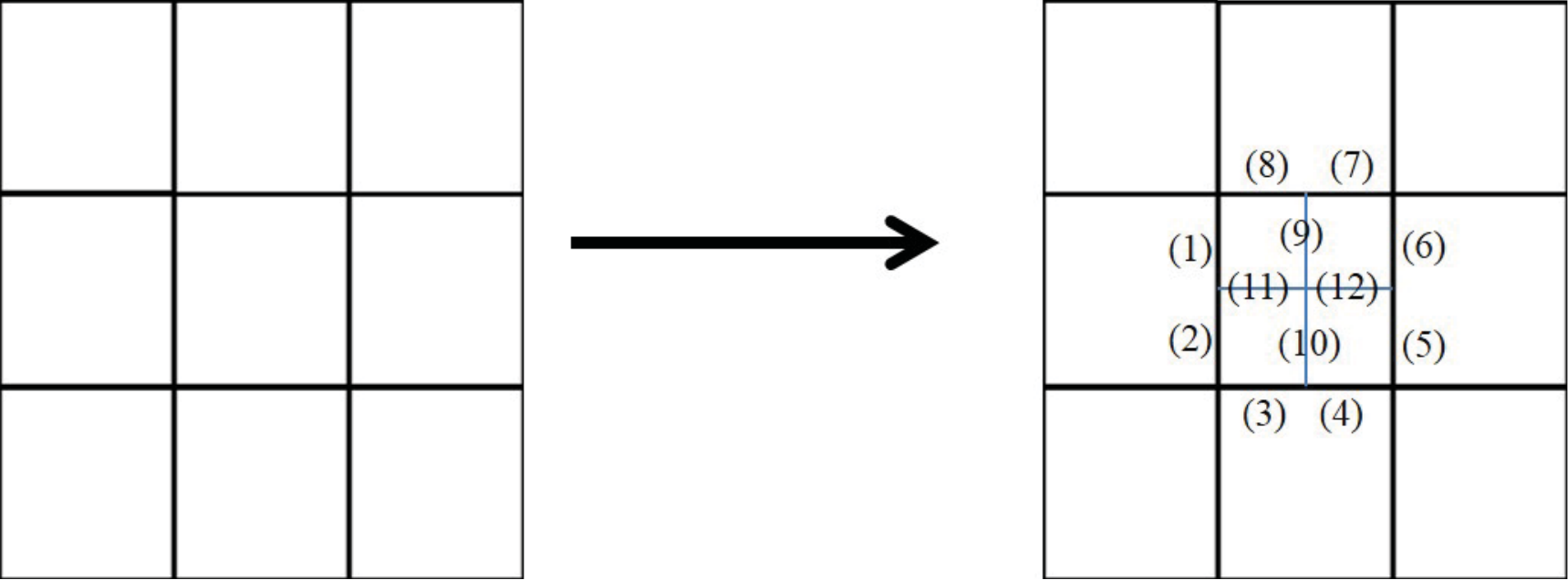}
\caption{Schematic diagram for the refinement operation acting on the magnetic field in the two-dimensional example. The numbers stand for the values on the faces to be determined by the Balsara's method described in the text.}
\label{fig: AMR_prolongation}
\end{figure}

Following GAMER, we define the base grid resolution as level $0$ and the $n$-th refinement as level $n$, where level $n$ has a spatial resolution twice higher than level $n-1$. Data in GAMER are always decomposed into {\it{grid patches}} (hereafter patches for short), each of which consists of $N^3$ cells, and the AMR implementation is realized by constructing a hierarchy of patches in an octree structure. The MHD implementation is similar to HD except for more equations. Two operations are required for updating the AMR data at the moment when the adjacent levels (say, level $l$ and $l+1$) are synchronized.

1. {\it{Correction Operation}}.   At this moment the coarse-grid data might be slightly inconsistent with the fine-grid data since they are evolved independently, and this is a procedure to correct the coarse-grid data by the fine-grid data.  Two different situations can arise and the correction procedures are different. First, for a coarse grid overlapping with fine grids, the coarse-grid data are simply replaced by the volume average and area average of the fine-grid data for conserved variables and the magnetic field, respectively.  Second, for a leaf coarse grid adjacent to a coarse-fine interface, the flux difference between the coarse and fine grids on the interface will be used to correct the coarse-grid conserved variables adjacent to this interface.  A similar procedure is applied on the magnetic field correction where the electric field differences on coarse-fine interedges are used to correct the magnetic field.  These corrections are necessary to preserve the conservation law of conserved variables and the divergence-free constraint on the magnetic field.

2. {\it{Refinement/Derefinement Operations}}. This is the procedure to create/remove fine grids according to the refinement criteria. In creating new fine grids, we adopt the conservative interpolation for conserved variables and the interpolation method proposed by \citet{Balsara2001} to maintain a divergence-free magnetic field, which is also implemented in, for example, the ENZO \citep*{Bryan2014} and CHARM \citep*{Miniati2011} codes. In the following we use a 2D example to explain the Balsara's method.

Consider a coarse grid that needs to be refined to fine grids as illustrated by Fig. (\ref{fig: AMR_prolongation}). There are $12$ magnetic field refined values to be determined. We divide these $12$ values into two sets. The first set includes those interfacing with the coarse grids (labeled as number (1)-(8)) and the second set includes the interior interfaces (number (9)-(12)).  Values in the first set are determined by the piecewise linear interpolation among nearby coarse grids, which ensures that the average of fine magnetic field values is the same as the coarse value. For determining the second set (the interior part), we Taylor-expand the magnetic field about the coarse grid center,
   \begin{equation}
   \label{equ: B Taylor expansion}
   \left.\begin{aligned}
   & B_x(x,y) = a_0 + a_x x + a_y y +a_{xx}x^2  + a_{xy} xy, \\
   & B_y(x,y) = b_0 + b_x x + b_y y +b_{yy}y^2  + b_{xy} xy,
   \end{aligned}
   \right.
   \end{equation}
with $10$ coefficients to be determined. Here we ignore $y^2$ term in $B_x$ and $x^2$ in $B_y$ to fit linear profiles on the coarse grid interface, which is consistent with the piecewise linear interpolation acted on the first set. Evaluating Eq. (\ref{equ: B Taylor expansion}) for the first set (interfaces of the coarse grid) gives eight equations. However, these eight equations are not independent due to the divergence-free constraint on the coarse-grid magnetic field. Hence only seven independent equations survive from the matching values on interfaces. On the other hand, applying the divergence-free constraint on Eq. (\ref{equ: B Taylor expansion}) gives $\partial_x B_x + \partial_y B_y = (a_x + b_y) + (2a_{xx}+b_{xy})x + (a_{xy}+2b_{yy})y=0$. Values in the parentheses should separately be equal to zero and it gives three equations. In the end, we have 10 equations to uniquely determine the 10 coefficients in Eq. (\ref{equ: B Taylor expansion}). Once fixing these coefficients, interior values can be determined by evaluating Eq. (\ref{equ: B Taylor expansion}) at any appropriate position.

The procedure of updating grids at level $l$ can be summarized as follows.

1. Update all quantities (conserved variables and the magnetic field) for all grids at level $l$.

2. Evolve the next refinement level $l+1$ until the data at levels $l$ and $l+1$ are synchronized in time.

3. Apply the correction operation to correct quantities at level $l$.

4. Apply the refinement/derefinement operations to allocate/deallocate grids at level $l+1$ according to the refinement criteria.

Note that GAMER supports adaptive time-step integration where higher levels can have smaller time-steps. See \citet{Schive2017} for details.

\section{Hybrid MPI/OpenMP/GPUs Parallelization}
\label{sec: GAMER MHD Structure}

\begin{figure*}
\centering
\includegraphics[scale=0.68]{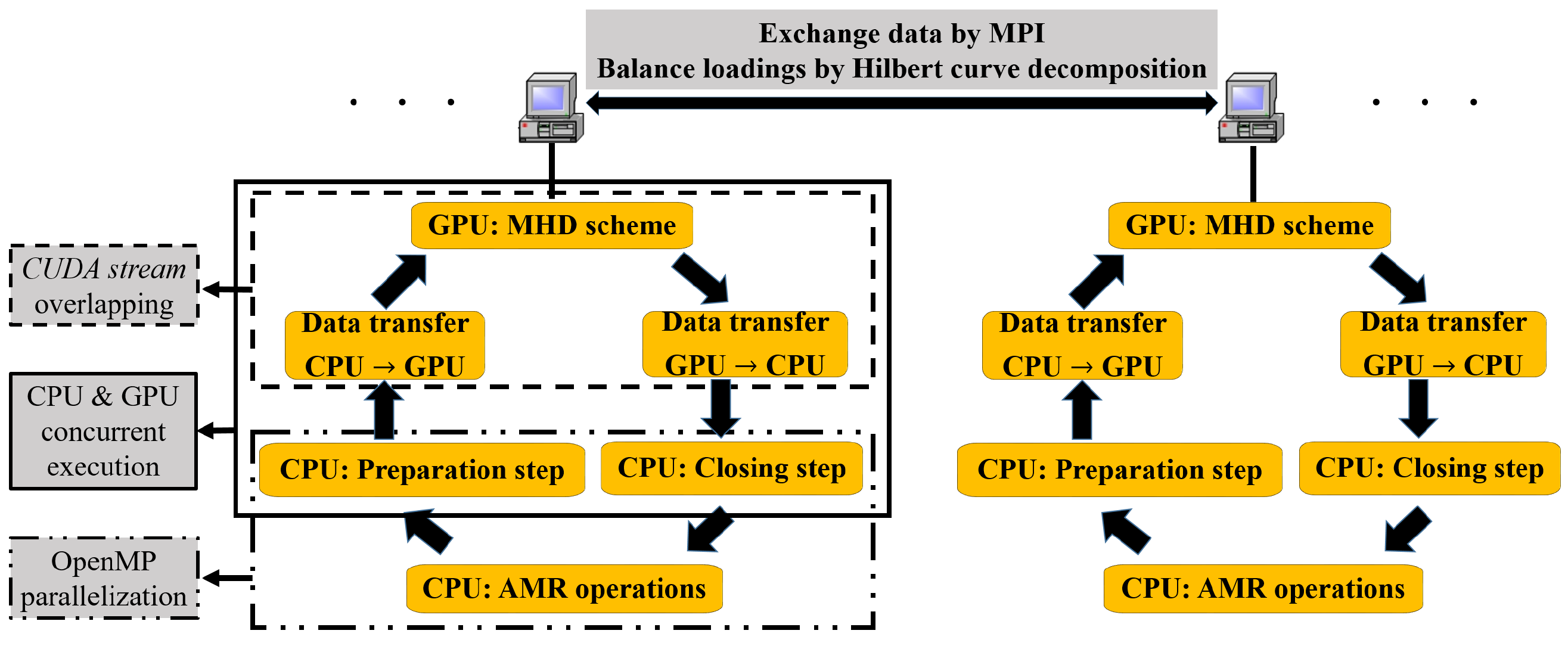}
\caption{Schematic diagram of hybrid MPI/OpenMP/GPUs parallelization in GAMER. Data among different computing nodes are exchanged by MPI and load balance is achieved by the Hilbert space-filling curve domain decomposition. Three optimization strategies (CUDA stream overlapping, CPU \& GPU concurrent execution and OpenMP parallelization) are highlighted with grey blocks on the left.}
\label{fig: GAMER_strucure_speed_up}
\end{figure*}

In GAMER, the MHD scheme mentioned in Sec. (\ref{subsec: MHD Scheme}) is executed on GPUs in parallel.  Nearby eight patches are grouped into a single {\it{patch group}} (which contains $(2N)^3$ cells), and each patch group is computed by one CUDA thread block. Choosing the patch group rather than a single patch as the computing unit can reduce the computational overhead associated with ghost zones due to the smaller surface/volume ratio. On the other hand, AMR operations e.g., the correction and refinement/derefinement operations mentioned in Sec. (\ref{subsec:Adaptive Mesh Refinement}) are executed on CPUs because they are much less time-consuming than the MHD scheme. GAMER stores all simulation data in the CPU memory, which is generally much larger than the GPU global memory, and sends only a small number of patch groups (typically a few hundreds) into GPU at a time.

GAMER has implemented several performance optimization strategies which can also be applied to MHD.  For example, the {\it{CUDA stream}} is used to overlap the MHD scheme execution in GPU with the data transfer between CPU and GPU. Concurrent execution between CPU and GPU hides the preparation and closing steps in CPU by GPU execution. The preparation step is for fetching appropriate data (patch group with its ghost zones) from the octree for the MHD scheme to update, and the closing step is for storing the newly updated data back to the octree. Note that ghost zones are only temporarily allocated for the patch groups being updated, which greatly reduces the memory consumption. Furthermore, Open Multi-Processing (OpenMP) is used to parallelize the preparation and closing steps and other AMR operations.

In addition, inter-node parallelization is achieved using Message-Passing-Interface (MPI), and Hilbert space-filling curve domain decomposition is applied for load balance. We compute Hilbert curves on different levels independently to achieve load balancing on a level-by-level basis. See \citet{Schive2017} for details.  This hybrid MPI/OpenMP/GPUs implementation allows for an efficient exploitation of the computing power of multi-node/multi-core CPU/multi-GPUs in heterogeneous supercomputers.  Figure (\ref{fig: GAMER_strucure_speed_up}) summarizes the schematics of this hybrid scheme. Data among different computing nodes are exchanged by MPI and load balancing is achieved by Hilbert space-filling curve domain decomposition. In each computing node, we update the fluid and magnetic field by iterating over all patches on a given AMR level, which starts from copying data from the octree patch data structure (and interpolating the ghost-zone data, if necessary) in the preparation step, sending these data from CPU to GPU, applying MHD scheme, sending updated data from GPU to CPU, and applying the closing step to store these new data back to the octree structure. After updating all patches on a given AMR level, we apply the AMR operations mentioned in Sec. (\ref{subsec:Adaptive Mesh Refinement}). We also highlight the optimization strategies in Fig. (\ref{fig: GAMER_strucure_speed_up}) by blocks on the far left.  More details can be found in \citet{bSc1}.

\section{numerical results}
\label{sec: numerical results}

In this section, we perform two different kinds of tests. The first is accuracy tests which show the correctness of GAMER-MHD code, and the second is performance tests that quantify the speed-up and scalability. In all tests the adiabatic index $\gamma$ is chosen to be $5/3$ and the Courant safety factor is set to $1/2$.

\subsection{accuracy test}
\label{subsec: accuracy test}

\bigskip

(\rmnum{1}) Linear wave test

\medskip

\begin{figure}
\includegraphics[scale=0.355]{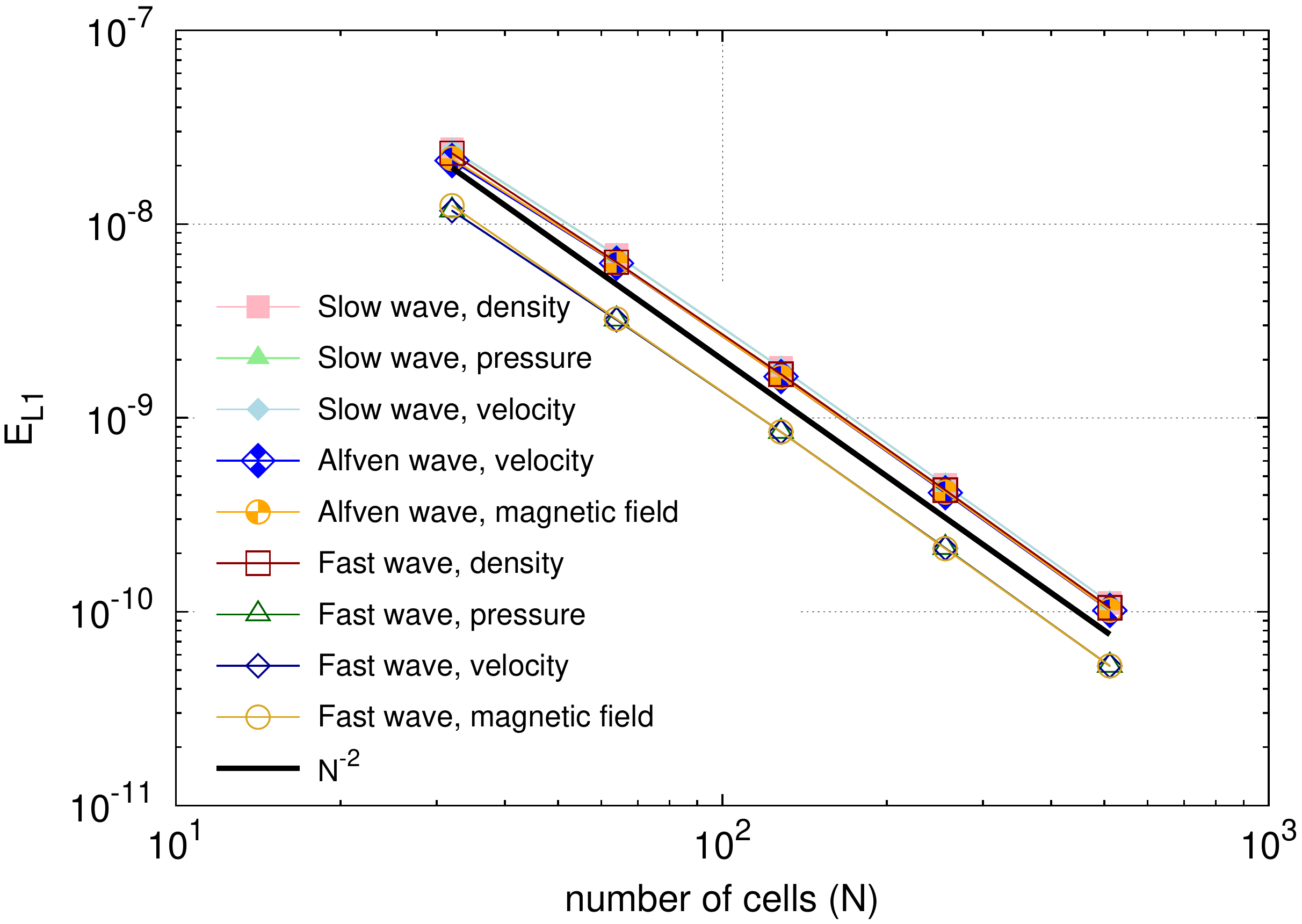}
\caption{L1-errors for the slow, Alfven and fast waves. These errors follow the inverse square of the cell number, implying a second order accuracy.}
\label{fig:liner_wave_test}
\end{figure}

There are three types of linear waves: slow, Alfven and fast waves. In principle, these waves are one-dimensional problems. However, we can make propagating velocity along the diagonal direction of the simulation box subject to the periodic boundary to examine the 3D MHD algorithm.

To quantify the accuracy, the traveling wave with the relative amplitude $1.0 \times 10^{-6}$ by uniform grids is performed for one period and the L1-error is measured. Here L1-error, $E_{L1}$, is defined as $E_{L1} \equiv \sum_{i=1}^N |u_i^T-u_i^0|/N$, where $N$ is number of cells and $u$ is primitive variables (density, velocity, pressure and magnetic field). The subscript stands for the grid index and superscript is for temporal label with $0$ for the initial time and $T$ for the one period. Figure (\ref{fig:liner_wave_test}) depicts our results. Here PLM data reconstruction, Roe Riemann solver, and the double precision calculation are adopted in these tests.

From Fig. (\ref{fig:liner_wave_test}), all errors are inversely proportional to the square of the number cells, consistent with the second order accuracy of the CTU scheme with the PLM data reconstruction.

\bigskip

(\rmnum{2}) Shock tube test

\medskip

\begin{table*}
\centering
\label{Table: L-R states for Riemann problems}
\caption{Left- and right-states for 1D Riemann Problems at $t=0$}{
\begin{tabular}{ccccccccccccccc} \hline \hline \\
Test
& $\rho_{L}$ & $V_{x,L}$ & $V_{y,L}$ & $V_{z,L}$ & $P_{L}$ & $B_{y,L}$ & $B_{z,L}$
& $\rho_{R}$ & $V_{x,R}$ & $V_{y,R}$ & $V_{z,R}$ & $P_{R}$ & $B_{y,R}$ & $B_{z,R}$
\\ \hline \\
Torrilhon & 1.0 & 0 & 0 & 0 & 1.0 & 1.0 & 0 & 0.2 & 0 & 0 & 0 & 0.2 & $\cos(3)$ & $\sin(3)$ \\ \hline \\
RJ2a & 1.08 & 1.2 & 0.01 & 0.5 & 0.95 & 3.6/$(4\pi)^{1/2}$ & 2/$(4\pi)^{1/2}$   &  1 & 0 & 0 & 0 & 1 & 4/$(4\pi)^{1/2}$ & 2/$(4\pi)^{1/2}$ \\ \hline \\
\end{tabular}
}
\end{table*}

\begin{figure*}
\centering
\includegraphics[scale=0.74]{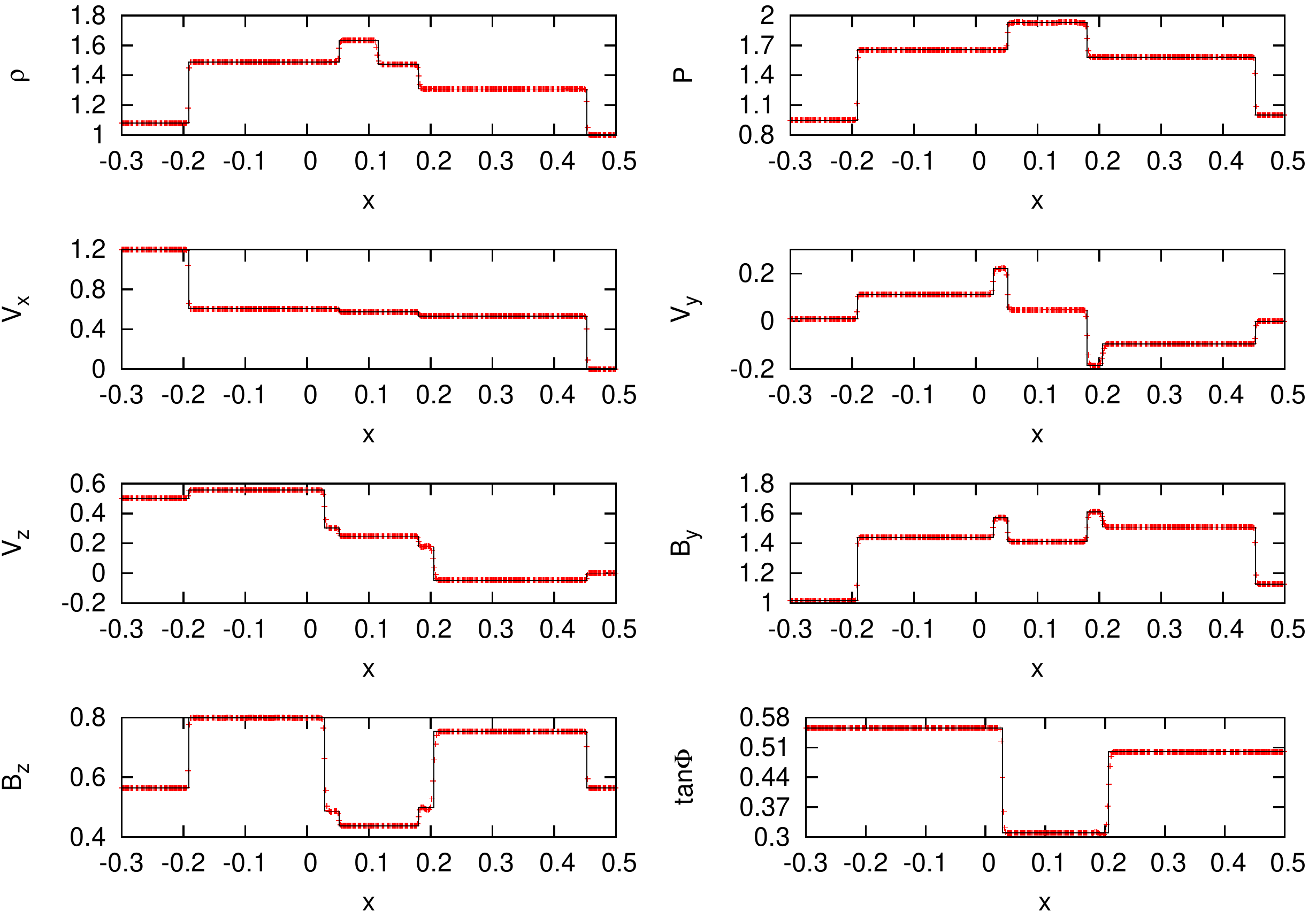}
\caption{Density, pressure, velocity components, transverse components of magnetic field, and the rotational angle $\Phi \equiv \tan^{-1}(B_z/B_y)$ for the RJ2a problem at $t = 0.2$. The simulation adopts 512 uniform grids with PPM data reconstruction and the Roe solver.  The exact regular solution \citep*{Torrilhon2002} is solid lines, and the simulation data are in red dots. All discontinuous waves are captured by $2-4$ cells.}
\label{fig: Riemanian_problem_RJ2a}
\end{figure*}
\begin{figure*}
\centering
\includegraphics[scale=0.74]{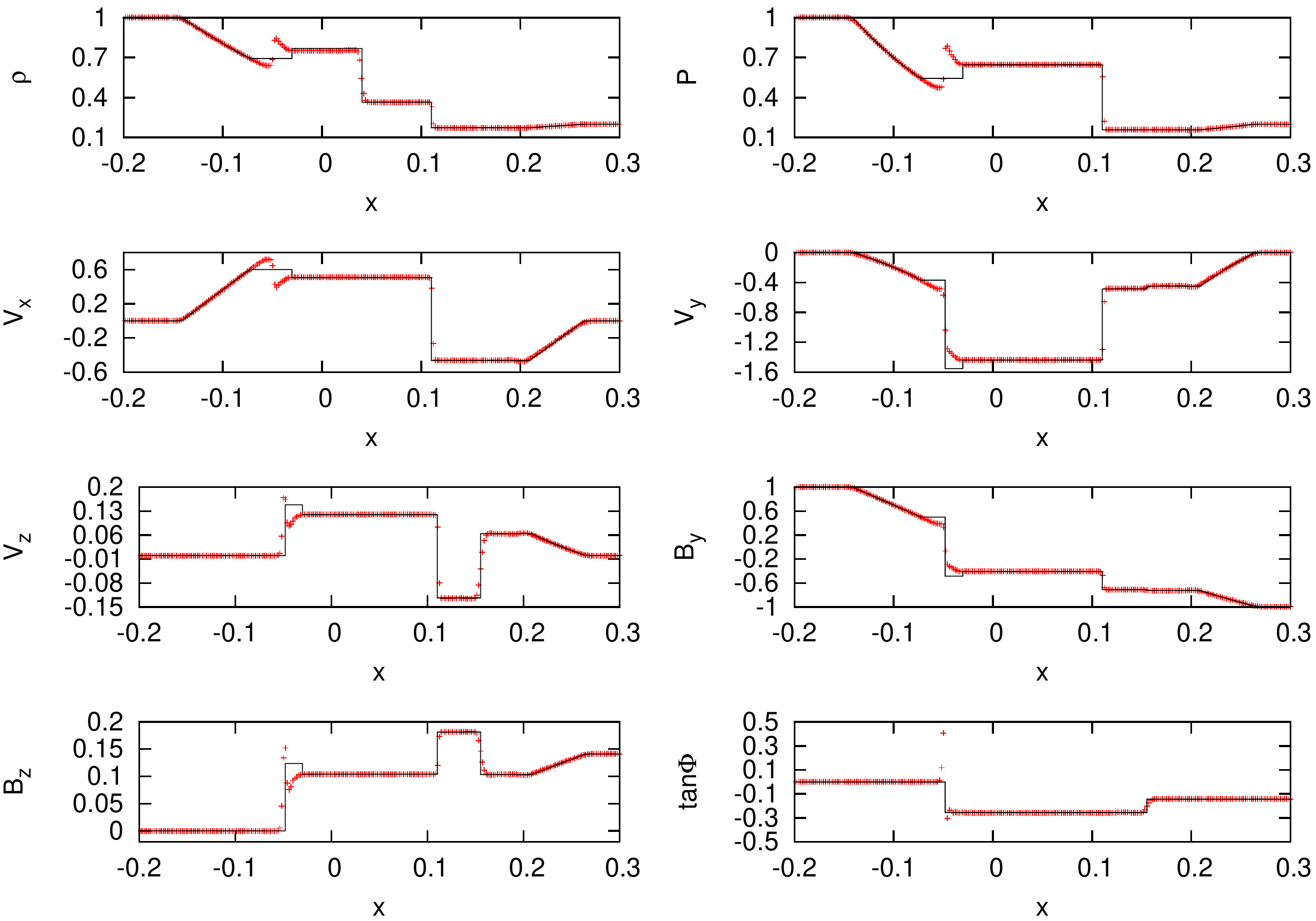}
\caption{Same as Fig. (\ref{fig: Riemanian_problem_RJ2a}) for the Torrilhon problem at $t=0.08$. The simulation setup is the same as the RJ2a test. Solid lines are the exact regular solution \citep*{Torrilhon2002}. The numerical solution deviates from the exact solution at the compound wave appearing near $x=-0.05$. This compound wave can be eliminated by increasing the spatial resolution but only at a very slow and unspecified rate.}
\label{fig: Riemanian_problem_Torrilhon}
\end{figure*}
\begin{figure*}
\centering
\includegraphics[scale=0.74]{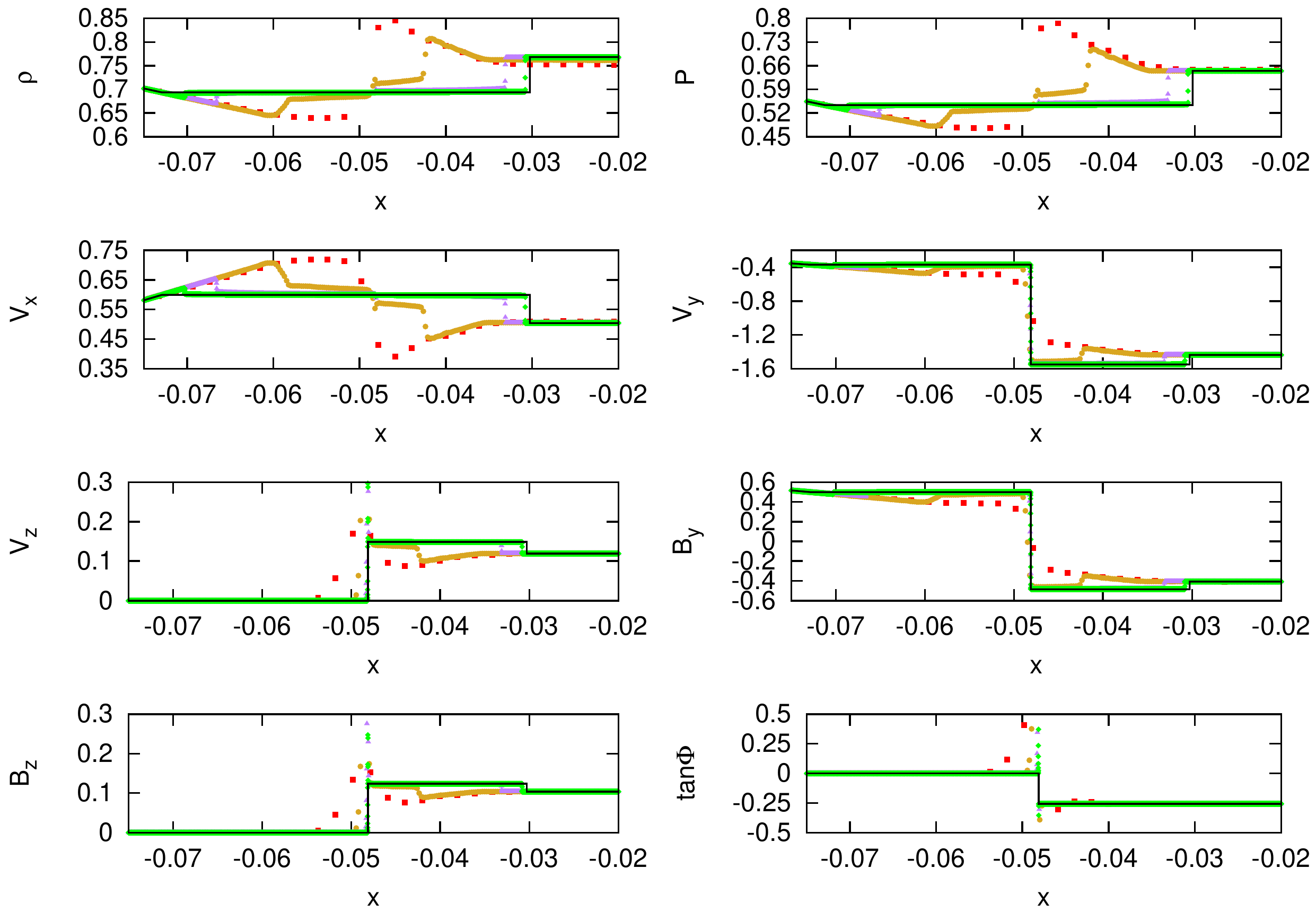}
\caption{Convergence tests for Torrilhon problem at $t = 0.08$. Notations are the same as Fig. (\ref{fig: Riemanian_problem_Torrilhon}).  Four results of different resolutions are shown and they are $2^9$ (filled square), $2^{12}$ (filled circle), $2^{15}$ (filled triangle) and $2^{18}$ (filled diamond) uniform grids, respectively. Also plotted here is the exact regular solution \citep*{Torrilhon2002} in solid lines. The numerical solution is seen to slowly approach the exact solution as the resolution increases.}
\label{fig: Riemanian_problem_Torrilhon_converge}
\end{figure*}
\begin{figure}
\includegraphics[scale=0.355]{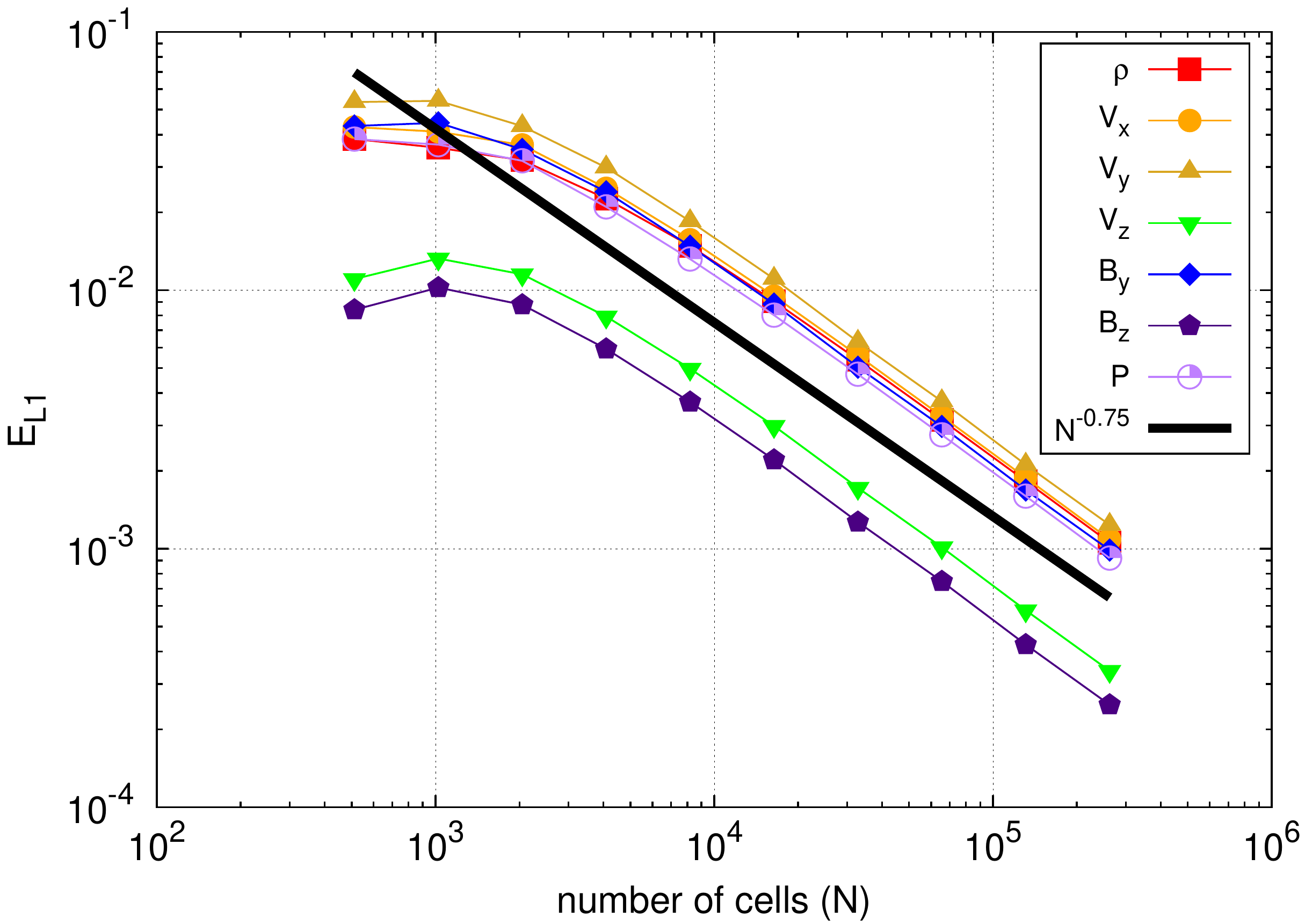}
\caption{L1-errors for the Torrilhon problem at $t = 0.08$. Notations are the same as Fig. (\ref{fig: Riemanian_problem_Torrilhon}).  The slow convergent rate can be fitted by the power law with a power index $-3/4$.}
\label{fig: Riemanian_problem_Torrilhon_L1_Error}
\end{figure}

Shock tube tests are for examining the goodness of simulating nonlinear waves, especially capturing shocks and discontinuities. Such kind of tests can be generated by one dimensional Riemann problems. Without losing of generality, we set the propagating direction along the $x$-axis, i.e., all variables depend only on $x$. We choose the \citet{Ryu1995} (RJ2a for short) and \citet{Torrilhon2003} Riemann problems for our tests. These two problems are also standard tests in the ATHENA code \citep*{Stone2008}. Table ($1$) is the initial left and right states for these two tests. Here PPM data reconstruction, Roe Riemann solver and the double precision calculations are adopted, and the boundary condition is Dirichlet type determined by the left and right states. The simulation box length is chosen to be $1$.

Figures (\ref{fig: Riemanian_problem_RJ2a}) and (\ref{fig: Riemanian_problem_Torrilhon}) are results. For MHD Riemann problems, the longitudinal magnetic field component ($B_x$) should be specified. Here  $B_x=2/(4\pi)^{1/2}$ for RJ2a and $B_x=1.0$ for Torrilhon. We simulate both cases with 512 uniform grids, which can be compared to the ATHENA code directly as this number of grids is also adopted in \citet{Stone2008}. All discontinuous waves are captured by $2-4$ cells, especially shown by the solution of the RJ2a problem where all discontinuities in each MHD wave family are reproduced, i.e., left- and right-propagating fast and slow shocks, left- and right-propagating rotational discontinuities and a contact discontinuity.  Small overshoot oscillations appear in the velocity and the magnetic field (with a relative amplitude of less than $0.1\%$ and thus not visible in Fig. (\ref{fig: Riemanian_problem_RJ2a})) and they can be eliminated by using PLM data reconstruction.

However, an unphysical compound wave structure is seen in the Torrilhon test in Fig. (\ref{fig: Riemanian_problem_Torrilhon}) near $x=-0.05$, due to numerical dissipation.  This structure can be eliminated by increasing the spatial resolution but only at a very slow and unspecified rate \citep*{Torrilhon2003}.  As this problem is intrinsically 1-D, we can employ a very high resolution uniform-grid simulation along the direction of propagation within a reasonable run time, by which the unspecified convergent rate can be found.  Figure (\ref{fig: Riemanian_problem_Torrilhon_converge}) is the result. The strength of the compound wave decreases and the slow shock on the right and the rotational discontinuity on the left are seen to get separated when the resolution becomes very high. There is an overshooting at the rotational discontinuity in the $z$-component of the velocity and magnetic field's profiles and the width of this overshooting shrinks as the resolution increases. Furthermore, we also measure the $E_{L1}$ of all quantities. The definition of $E_{L1}$ is the same as the linear wave test except for measuring the difference between the numerical and exact solutions. We focus on the region of the compound wave within in $-0.073<x<0.04$ at $t=0.08$, and measure the difference between the numerical solution and the exact solution in that region. Figure (\ref{fig: Riemanian_problem_Torrilhon_L1_Error}) presents the L1-Error versus the number of cells. The convergent rate is much slower than normal second order schemes and can be fitted as a power law with power index $-3/4$.

\bigskip

(\rmnum{3}) Orszag-Tang Vortex test

\medskip

\begin{figure*}
\centering
\includegraphics[scale=0.93]{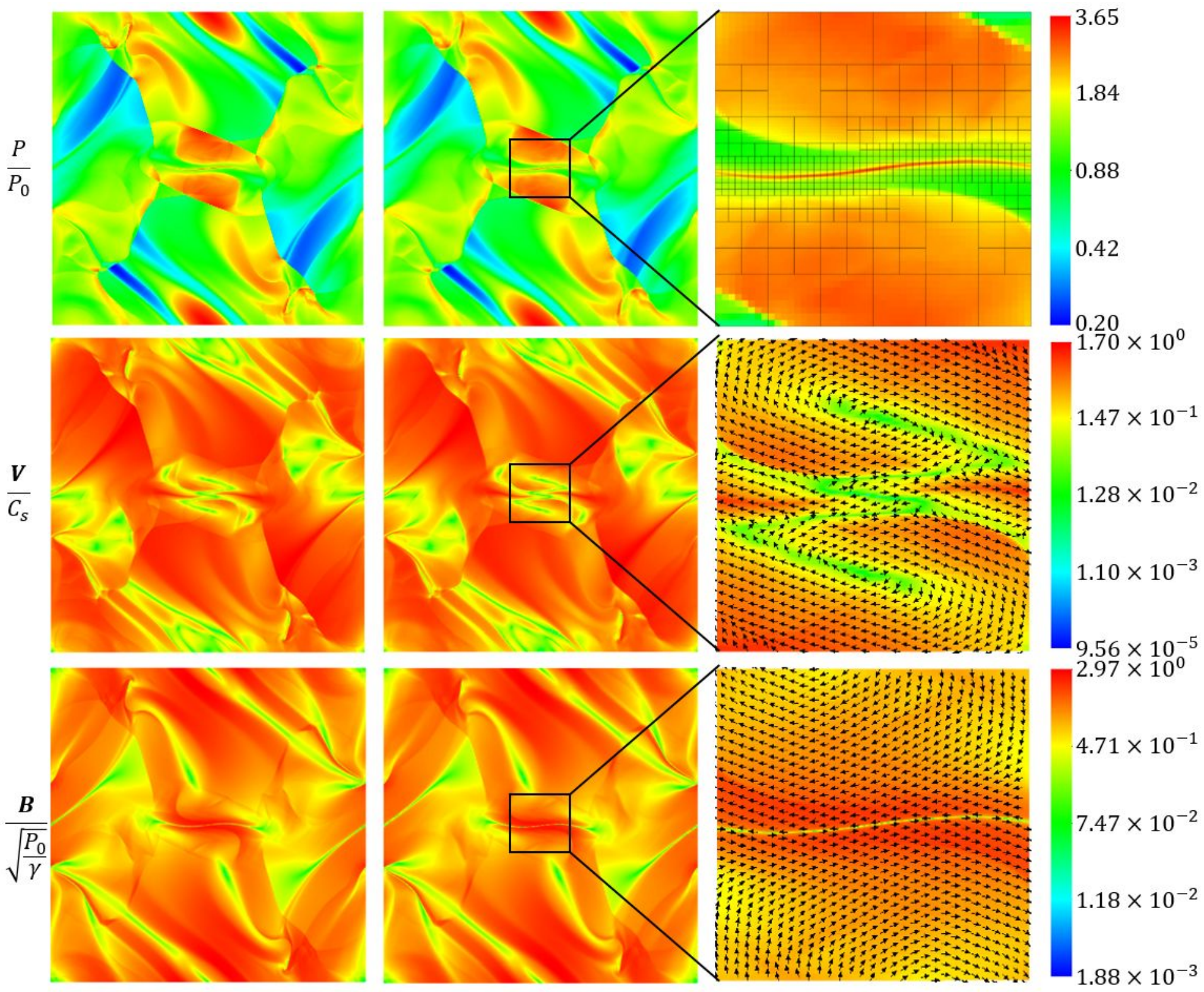}
\caption{The $xy$ images at $t=0.5L/C_s$ in the Orszag-Tang Vortex test. The left column is the $1024^2$ uniform-grid result and the middle column AMR result with $256^2$ base level grids and $3$ refinement levels. The right column is zoom-in images of the AMR result. The three rows present the pressure (normalized by $P_0$), velocity field (normalized by $C_s$) and magnetic field (normalized by $\sqrt{P_0/\gamma}$) from top to bottom, respectively. The color stands for the magnitude and arrows in velocity and magnetic fields the directions. The patch structure of AMR result is also shown with squares representing patches (consisting of $8^3$ cells) in the pressure zoom-in image. The AMR result agrees with the uniform case and the refined patches capture magnetic reconnection well, which is found similar to the Sweet-Parker type.}
\label{fig: Orszag_Tang}
\end{figure*}
\begin{figure*}
\centering
\includegraphics[scale=0.65]{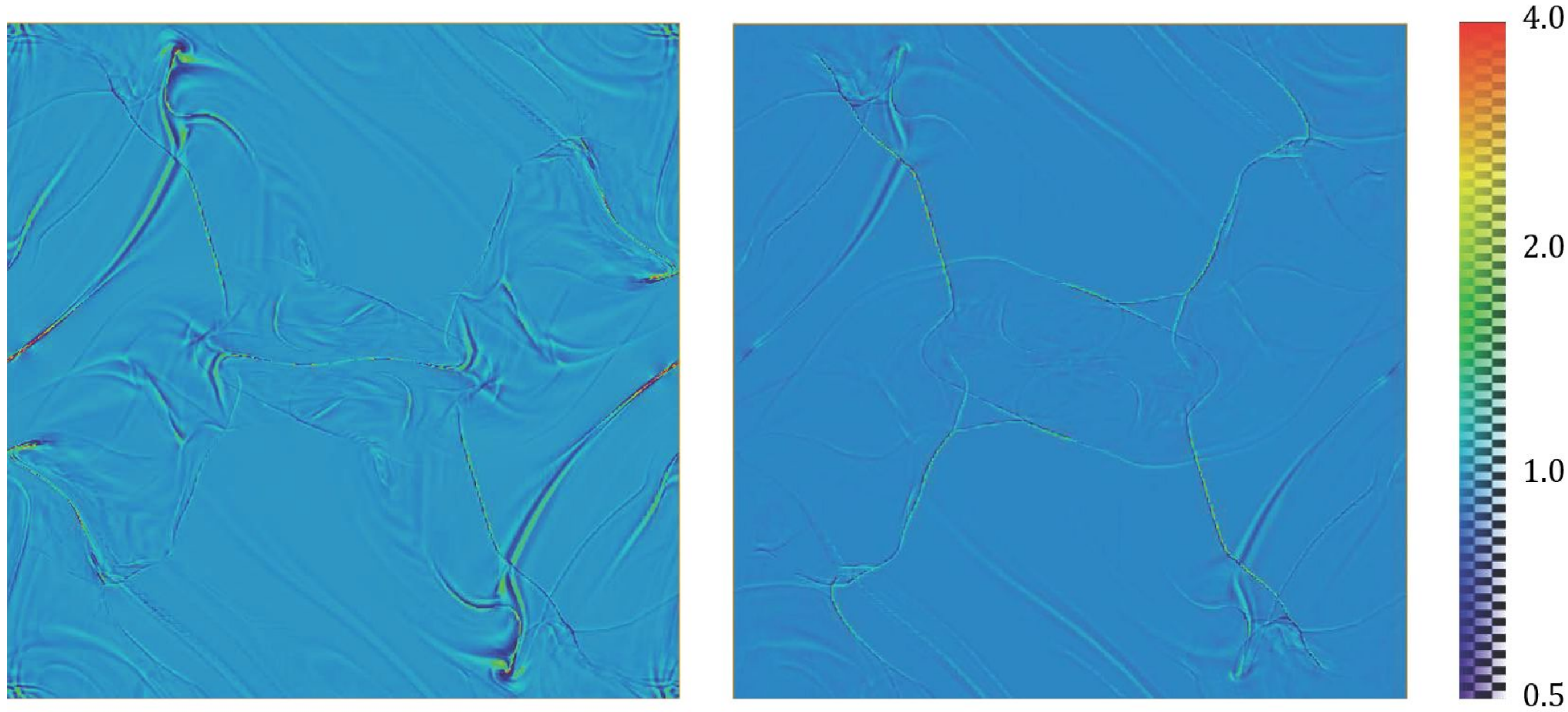}
\caption{Ratios between AMR and uniform-grid simulations for the magnetic pressure (left panel) and the gas pressure (right panel) at $t=0.5L/C_s$ in the Orszag-Tang Vortex test. The ratios are almost unity throughout the whole domain except for some filamentary regions corresponding to shock and reconnection locations.}
\label{fig: Orszag_Tang_quantative_anaylsis}
\end{figure*}

This test was developed by \citet{Orszag1979}. In this test, small-scale structure can be generated via a single large-scale differentially rotating vortex acting on two circular magnetic structures of opposite signs.  This is a two dimensional problem, so we start the simulation with a square box subject to the periodic boundary. The initial density $\rho_0$ and pressure $P_0$ are constants everywhere. The initial velocity and the magnetic field are given by
\begin{equation}
\label{equ: initial condition of v and B}
\left.\begin{aligned}
& \bm{V} = -C_s\sin \Big ( {{2\pi}\over L} y \Big ) \hat x + C_s\sin\Big ( {{2\pi}\over L} x \Big ) \hat y, \\
& \bm{B} = -\sqrt{{P_0}\over \gamma} \sin \Big ( {{2\pi}\over L} y \Big ) \hat x + \sqrt{{P_0}\over \gamma} \sin \Big ( {{4\pi}\over L} x \Big ) \hat y,
\end{aligned}
\right.
\end{equation}
where $C_s \equiv \sqrt{\gamma P_0/\rho_0}$ is the sound speed and $L$ the size of the box. The numerical setup (data reconstruction (PLM or PPM), Riemann solver and computation precision) is the same with shock tube tests.  In addition, we turn on AMR.

Figure (\ref{fig: Orszag_Tang}) depicts the results at $t=0.5L/C_s$. The left column in Fig. (\ref{fig: Orszag_Tang}) is made by the $1024^2$ uniform-grid simulation and the other two columns by the AMR simulation with $256^2$ base level grids and $3$ levels of refinement. The uniform-grid simulation has an equivalent resolution to the second refinement level in the AMR simulation. The quantity $|\bm{J}|/|\bm{B}|$ is used as the refinement criterion.  When the local $|\bm{J}|/|\bm{B}| > 3.2/\Delta h(k)$, the grid is refined to level $k+1$ where $\Delta h(k)$ is the $k$-th level grid size. This criterion aims to capture magnetic reconnection since $|\bm{J}|/|\bm{B}|$ becomes singular at reconnection sites. In the AMR case, there are about $3.6 \times 10^5$ grid cells and the volume-filling fractions on different levels are $24.0\%$, $8.5\%$ and $3.3\%$ from the lowest to the highest refinement levels, respectively, at $t=0.5L/C_s$.

Figure (\ref{fig: Orszag_Tang}) shows good match between the uniform-grid simulation and the AMR one. The quantitative analysis are in Fig. (\ref{fig: Orszag_Tang_quantative_anaylsis}), in which we measure the ratio between the data of the AMR result and the uniform-grid result for the magnetic pressure and gas pressure. AMR data are interpolated (averaged) to have the same spatial resolution with the uniform-grid data. Values in Fig. (\ref{fig: Orszag_Tang_quantative_anaylsis}) are almost unity except for at shock locations and the magnetic reconnection layer, indicative of good match between the AMR solution and the uniform-grid solution. The patch structures shown in the right column of Fig. (\ref{fig: Orszag_Tang}) illustrates that the refinement criterion captures magnetic reconnection quite well. Furthermore, reconnection found in this test appears to be a variant of Sweet-Parker reconnection \citep{Parker1957, Sweet1958} with two spiral inflows and two horizon jet outflows located near the mid plane as depicted in the zoom-in image in Fig. (\ref{fig: Orszag_Tang}). The inflow velocity structure has been modified by the large-scale differential rotation.

\bigskip

(\rmnum{4}) Blast wave test

\medskip

\begin{figure*}
\centering
\includegraphics[scale=0.83]{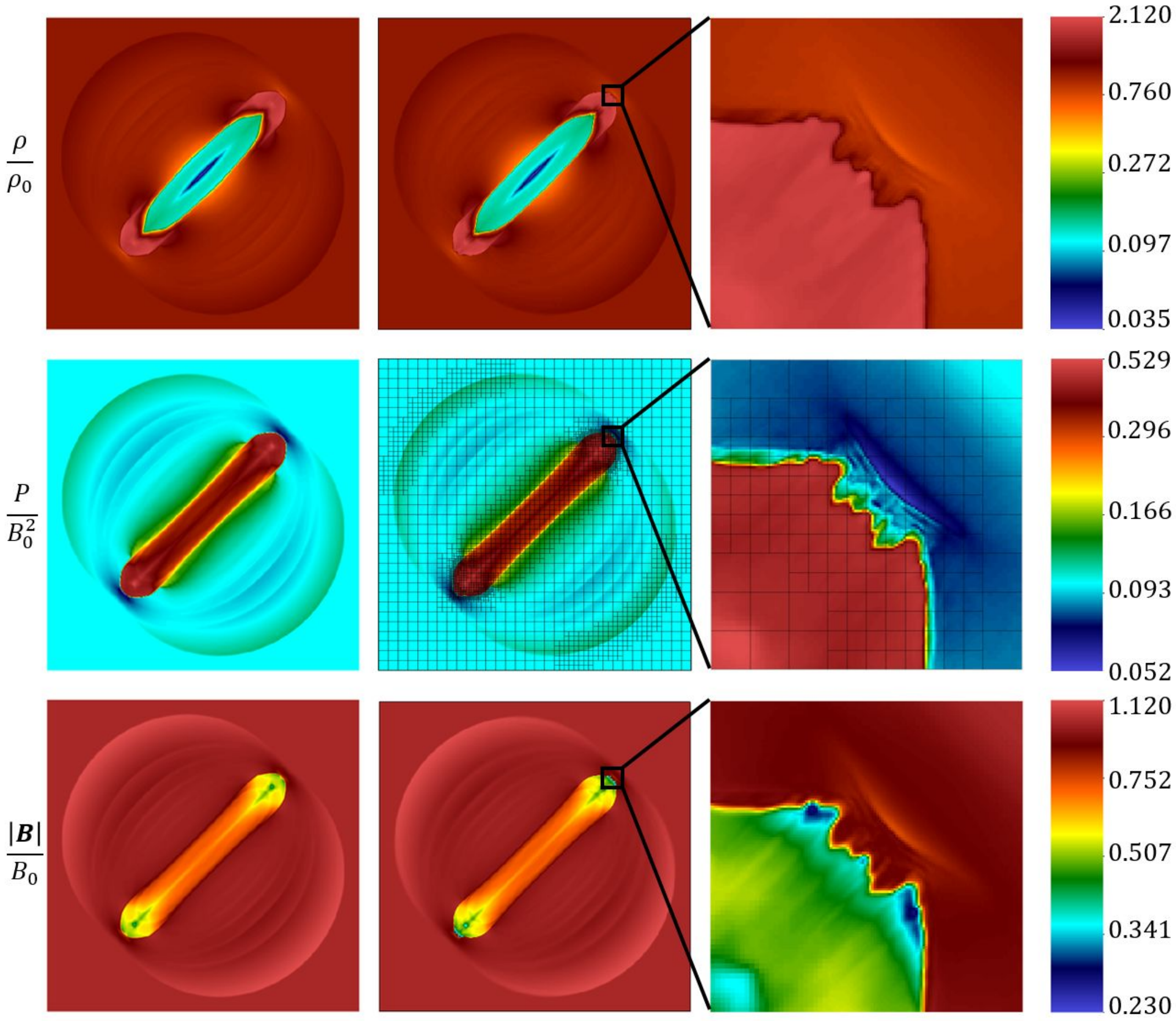}
\caption{The $xy$ slice images which passes through the box center at $t=6r_0/C_A$ in the blast wave test. The left column is the $512^3$ uniform grid result and the middle is AMR result with a base level $256^3$ up to level $3$. Quantities from top to bottom are the density (normalized by $\rho_0$), pressure (normalized by $B_0^2$) and magnetic field magnitude (normalized by $B_0$), respectively. We also depict the patch structure of AMR with one square representing a patch (consisting of $8^3$ cells here) in the pressure image. The refined patches capture the shocks. The AMR result has the "finger" structure in the shock nose along the magnetic axis (right column).}
\label{fig: Blast_wave}
\end{figure*}
\begin{figure*}
\centering
\includegraphics[scale=0.73]{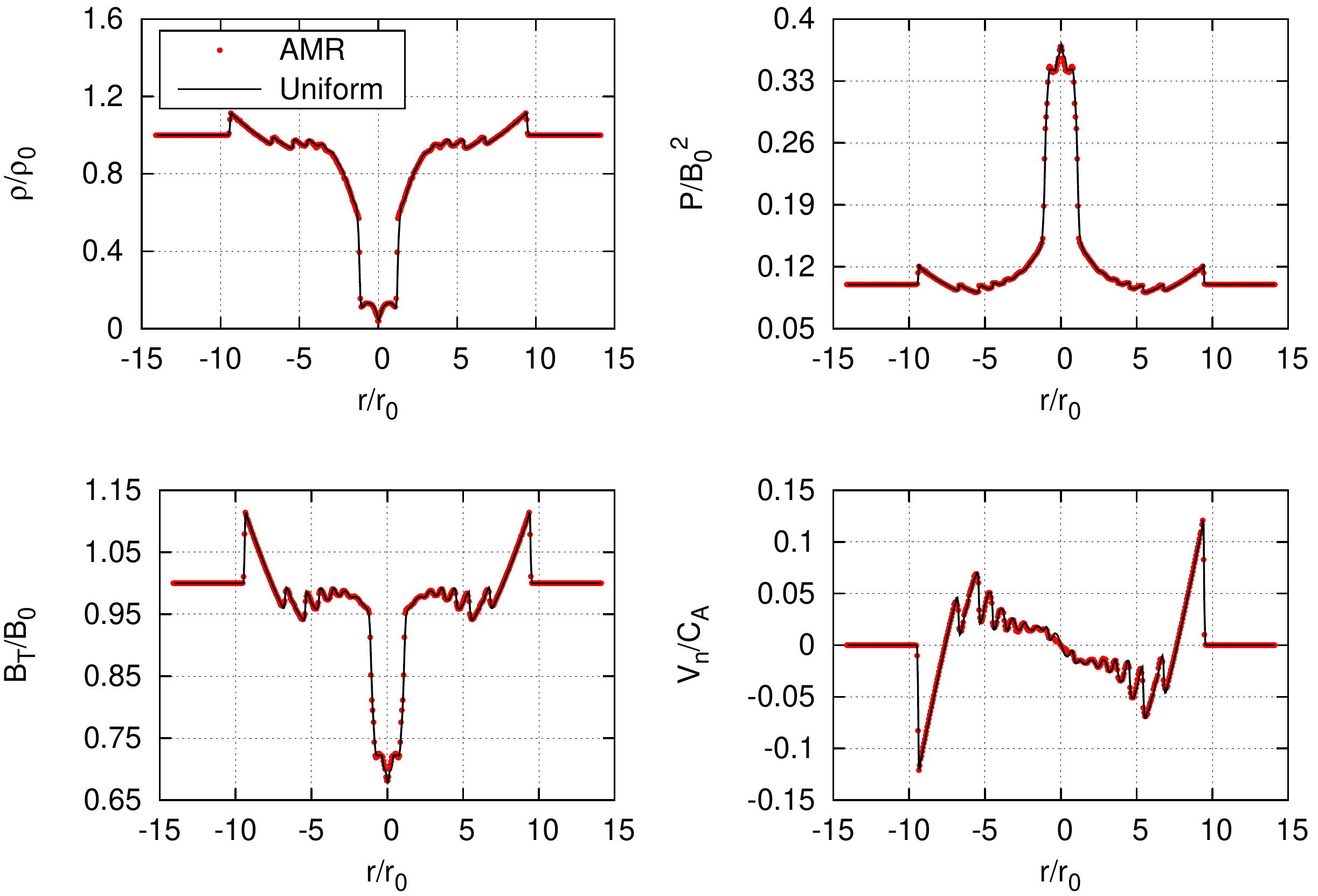}
\caption{Profile of a through-line to the box center on the equatorial plane at $t=6r_0/C_A$ in the blast wave test. Points present the AMR result and the solid line is the  $512^3$ uniform-grid result, respectively. Four panels show density (top left), pressure (top right), transverse component of the magnetic field (bottom left), and normal component of the velocity field (bottom right). The normal component of the magnetic field and the transverse component of the velocity field are negligibly small. The shocks located at $\pm9.4r_0$ are captured by $1$ to $2$ cells and the contact discontinuities located at $\pm 1.3r_0$ by $4-5$ cells. All shocks are fast shocks. The profiles of AMR match those of the uniform grid very well.}
\label{fig: Line_profile_equator}
\end{figure*}
\begin{figure*}
\centering
\includegraphics[scale=0.73]{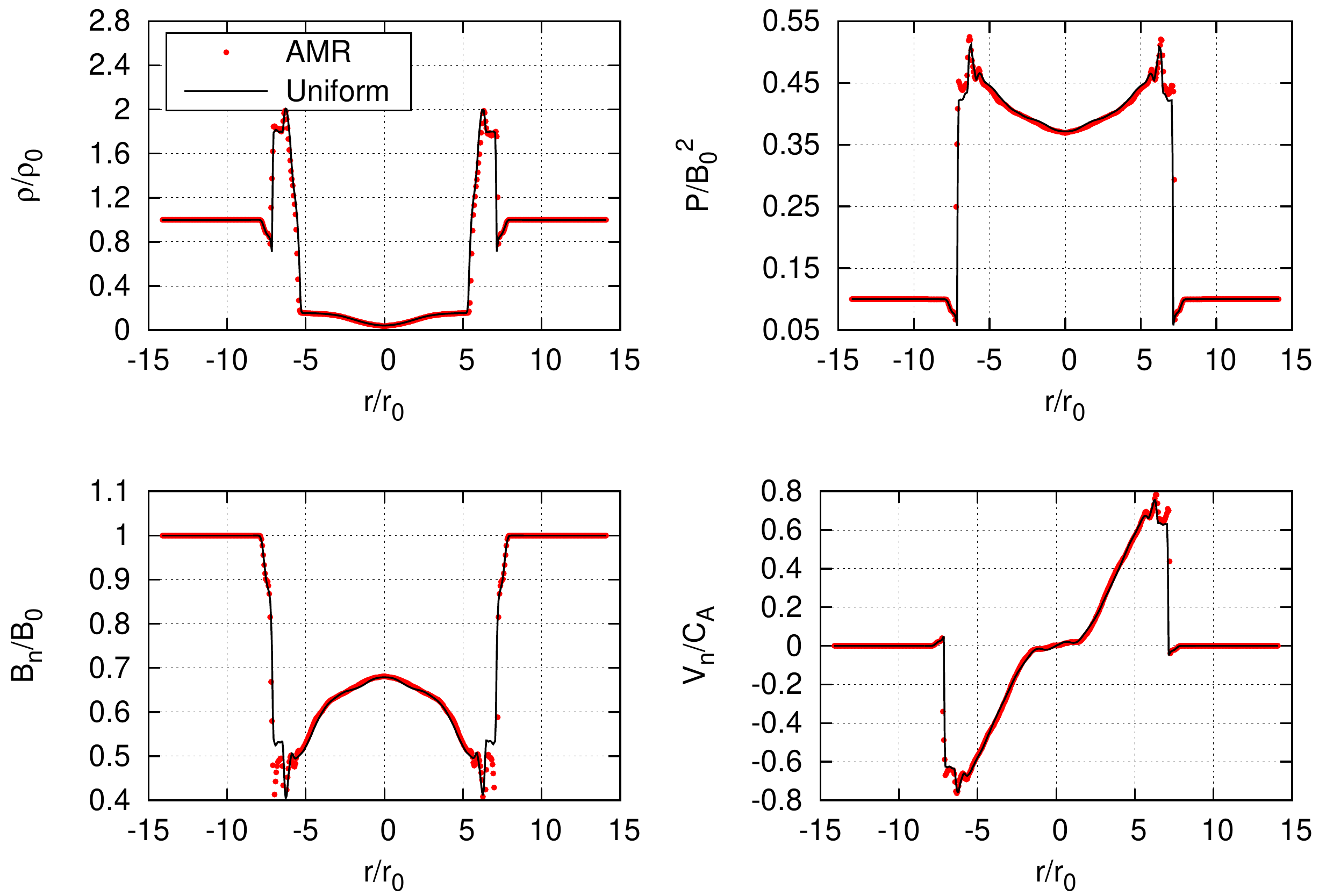}
\caption{Through-line profile along the symmetric axis at $t=6r_0/C_A$ in the blast wave test. Data are same as Fig. (\ref{fig: Line_profile_equator}), except for $B_T$ is replaced by $B_n$. The contact discontinuities are located at $r=\pm5.4r_0$ and mixed with the rarefaction wave. The shocks, a slow shock, are located at $r=\pm7.2r_0$. The AMR result agrees with the uniform-grid case except at the downstream of the slow shock located near $r=\pm 7r_0$. The deviation is from the ``finger'' structure appearing on the nose of the slow shock in the AMR case but not in the uniform-grid case.}
\label{fig: Line_profile_symmetric_axis}
\end{figure*}
\begin{figure*}
\centering
\includegraphics[scale=0.42]{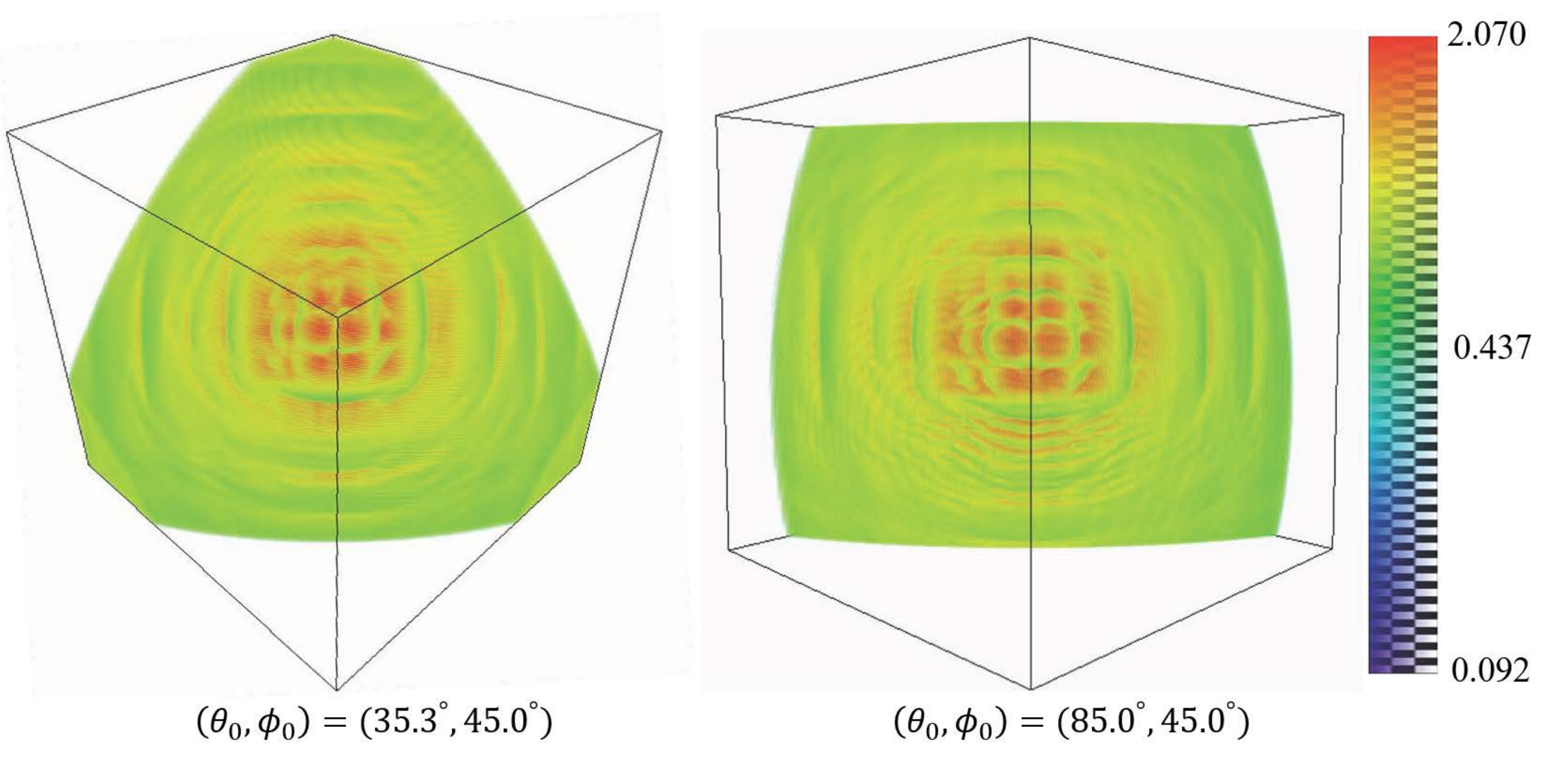}
\caption{Pressure slices (normalized by $B_0^2$) for two simulations of different directions of ambient magnetic fields, which are oriented to $(\theta_0, \phi_0) = (85.0^{\circ}, 45.0^{\circ})$ on the right and $(\theta_0, \phi_0) = (35.3^{\circ}, 45.0^{\circ})$ on the left at $t=0.5r_0/C_A$ in the blast wave test, where the initial density perturbations are given in Eq. (\ref{equ: formula of the density perturbation}). The box size is the same as the size of the zoom-in image in Fig. (\ref{fig: Blast_wave}). The left and right small structures are similar, demonstrating the structure is real, except for the resolved ringing pattern in the lower middle region of the right panel not resolved in the left panel.}
\label{fig: Blast_wave_verify_stability}
\end{figure*}

This test examines the propagation of strong MHD shock in a magnetized plasma. We generalize the initial condition in \citet{Londrillo2000} to three dimensional setup, i.e., a uniform, static, spherical plasma with radius $r_0$, density $\rho_0$ and pressure $P_0$ is surrounded by an ambient medium with constant density $\rho_0$ and pressure $10^{-2}P_0$. The magnetic field is uniform everywhere with the strength $B_0$. Values of $r_0$, $\rho_0$, $P_0$ and $B_0$ follow \citet{Londrillo2000}. In particular, we set the ambient plasma $\beta=0.2$. To characterize the strength of explosion, the maximum fast shock Mach number is found to be $M \equiv S/\sqrt{C_A^2+C_s^2} =2.54$ perpendicular to the magnetic field, where $S$ is the fast shock speed, $C_A$ and $C_s$ are the upstream Alfven speed and sound speed, respectively. Numerical schemes are the same with Orszag-Tang test. Finally, the simulating domain is a cube with a size $20r_0$ and the boundary is periodic.

Although the initial setup is three dimensional, this configuration has the rotational symmetry with respect to the magnetic field line. Therefore, the initial magnetic field can be of any arbitrary direction. We choose the magnetic field along with the diagonal direction of $xy$ plane, i.e., $\bm{B} = (B_0/\sqrt{2}, B_0/\sqrt{2}, 0)$.

Figure (\ref{fig: Blast_wave}) shows the $xy$ slice image which cuts through the center of explosion and the blast wave noses at $t=6r_0/C_A$. In Fig. (\ref{fig: Blast_wave}), the left column is made with the simulation with uniform $512^3$ cubic grids and the other two columns with the AMR simulation with a $256^3$ base level and three refinement levels. The uniform grid simulation has the equivalent resolution to the first refinement level in the AMR simulation.  The AMR refinement criterion is as follows. The grid with normalized pressure gradient ($10^{2}r_0|\nabla P|/P$) higher than $1.25 \times 2^{n-1}$ is refined to level $n$.  This condition aims to capture the strong shock. The AMR case has about $8.3 \times 10^7$ grid cells and the volume-filling fractions from the lowest to the highest refinement levels are $27.3\%$, $0.7\%$ and $0.3\%$, respectively, at $t=6r_0/C_A$.

From Fig. (\ref{fig: Blast_wave}), the AMR simulation agrees with the uniform-grid simulation on the large-scale structure. To see how well our scheme captures three dimensional shocks and discontinuity, we show profiles of lines penetrating through the equator and along the symmetric axis in Figures (\ref{fig: Line_profile_equator}) and (\ref{fig: Line_profile_symmetric_axis}), respectively. The profile on the equatorial plane reveals shocks are captured by $1$ to $2$ cells and the contact discontinuity by $4$ to $5$ cells. Weak shocks are produced in the region between the contact discontinuity and the strong shock. They arise from the oscillating contact discontinuity emitting a wave train, which is steepen to become shocks. Shocks on the equatorial plane are fast shocks.

The profile of a line along the symmetric axis is showed in Fig. (\ref{fig: Line_profile_symmetric_axis}). The contact discontinuity is located at about $r = \pm5.4r_0$ and captured by $4-5$ cells.  The contact discontinuity is mixed with a steep rarefaction wave, making it difficult to identify. The shock, located at about $r=\pm7.2r_0$, is captured by $1$-$2$ cells and it is a slow shock.  However, this slow shock is not the same as the usual one-dimensional slow shock, for which only the ``transverse component'' of the magnetic field changes across the shock.  Instead, we find the ``normal component'' of the magnetic field changes across the shock (c.f. Fig. (\ref{fig: Line_profile_symmetric_axis})). The suppression of the normal component downstream is due to the fact that upstream magnetic field lines suddenly fan out away from the axis across the shock, and hence the normal component of the magnetic field along the axis must decrease.  Finally, upstream of the slow shock is different from the ambient plasma because the oblique fast shock near the axis emits fast waves downstreams and influences the upstream region of the slow shock, which can be obviously seen in Figs. (\ref{fig: Blast_wave}) and (\ref{fig: Line_profile_symmetric_axis}) with a density depression on the axis immediately upstream of the slow shock.

We also plot in Fig. (\ref{fig: Line_profile_equator}) and (\ref{fig: Line_profile_symmetric_axis}) the uniform-grid result for comparison. The equatorial line profile agrees with the AMR result to a high accuracy. The axial line profile has slight deviations, notably near the slow shock. This brings us back to Fig. (\ref{fig: Blast_wave}), where a subtle feature is not captured in the uniform-grid simulation. The small-scale ``finger'' pattern appears at the slow shock nose near the axis, as shown in the right column in Fig. (\ref{fig: Blast_wave}), zoom-in images at the shock nose. To demonstrate this is a physical instead of numerical instability, we conduct another two simulations where the magnetic axis is rotated to two different orientations so that the grid geometry is different. In general, we can express the magnetic field as $\bm{B}/B_0=(\cos\phi_0\sin\theta_0, \sin\phi_0\sin\theta_0, \cos\theta_0)$ in the simulating box coordinate $(x, y, z)$. Here $\theta_0$ is the angle between $z$-axis and the magnetic field, and $\phi_0$ is the angle between $x$-axis and the projected magnetic field on $xy$ plane. In the following, we choose $(\theta_0, \phi_0) = (35.3^{\circ}, 45.0^{\circ})$, in which the magnetic field is along the diagonal direction of the simulating box, i.e, $\bm{B}/B_0=(1/\sqrt{3}, 1/\sqrt{3}, 1/\sqrt{3})$ and $(\theta_0, \phi_0) = (85.0^{\circ}, 45.0^{\circ})$, in which magnetic field is tilted by $5^{\circ}$ from the $xy$ plane and its $xy$-plane projection is aligned with the diagonal direction as our two different magnetic field orientations simulations. The $5^{\circ}$ inclining from $xy$ plane in the second case is to avoid Carbuncle instability, which is a numerical instability \citep*{Quirk1994}.

If no control noise is added to the system, the numerical errors of the second order accuracy in our scheme can produce different unstable patterns.  In the control test, we inject a short-wavelength density noise with no pressure fluctuation into a sphere of radius $1.5 r_0$, which is large enough to enclose the blast wave. To elaborate the perturbation pattern, we introduce a new coordinate $(x^{'}, y^{'}, z^{'})$ which can be specified by rotating the simulating box coordinate $(x, y, z)$ by the angle $\theta_0$ clockwise with respect to the rotational axis along the line $l:\{ x+y=0; z=0\}$. The angle $\theta_0$ is $35.3^{\circ}$ for the diagonal direction case and $85.0^{\circ}$ for the other. This rotation makes the initial magnetic field parallel to $z^{'}$-axis, i.e., $\bm{B} = B_0\hat z^{'}$, in each one of the simulation. The density perturbation in this new coordinate $(x^{'}, y^{'}, z^{'})$ has the following form,

\begin{equation}
\label{equ: formula of the density perturbation}
\left.\begin{aligned}
{{\delta \rho}\over{\rho_0}} = 2 \times 10^{-2} \Big [ & \sin \Big ( {{2\pi}\over{0.08r_0}}x^{'} + {{\pi}\over4}\Big ) + \sin \Big ( {{2\pi}\over{0.10r_0}}x^{'} \Big ) + \\
                                                       & \sin \Big ( {{2\pi}\over{0.12r_0}}x^{'} - {{\pi}\over4}\Big ) + \cos \Big ( {{2\pi}\over{0.08r_0}}y^{'} + {{\pi}\over4}\Big ) + \\
                                                       & \cos \Big ( {{2\pi}\over{0.10r_0}}y^{'} \Big ) + \cos \Big ( {{2\pi}\over{0.12r_0}}y^{'} - {{\pi}\over4}\Big ) \Big ].
\end{aligned}
\right.
\end{equation}
Therefore, runs of these two different orientations see the same noise pattern when this pattern is viewed from the magnetic axis.  For both reasons of demonstrating the reality of this finger structure and also capturing the wavelength of the perturbation, we adopt a static mesh refinement simulation.  The sphere that contains the short-wavelength noise is refined to resolve the shortest noise wavelength by $16$ cells.

Figure (\ref{fig: Blast_wave_verify_stability}) shows the pressure at $0.5r_0/C_A$ on the shock nose viewed from the magnetic axis for these two simulations. The figure shows largely similar patterns for these two simulations and eliminates the possibility of numerical instability. In detail, there are some very fine ``ringing'' pattern present in the case $(\theta_0, \phi_0) = (85.0^{\circ}, 45.0^{\circ})$ absent in the other. The separation of fine rings is about $2-3$ cells.  By contrast, the grid geometry for the other case ($(\theta_0, \phi_0) = (35.3^{\circ}, 45.0^{\circ})$) takes $\sqrt{3}$ grids to resolve structures in the direction of the ring pattern, and therefore becomes hardly able to resolve the rings.

This small-scale feature on shock noses likely arises from the dynamical instability similar to Rayleigh-Taylor type \citep*{Rayleigh1883, Taylor1950} in a decelerating shock.  Comprehensive analysis will be made in a forthcoming work.  At this stage, we simply show that the instability is indeed a physical one and can be captured by our high-resolution AMR simulation.

\bigskip

(\rmnum{5}) Magnetic field with the ABC pattern

\medskip

\begin{figure*}
\centering
\includegraphics[scale=0.80]{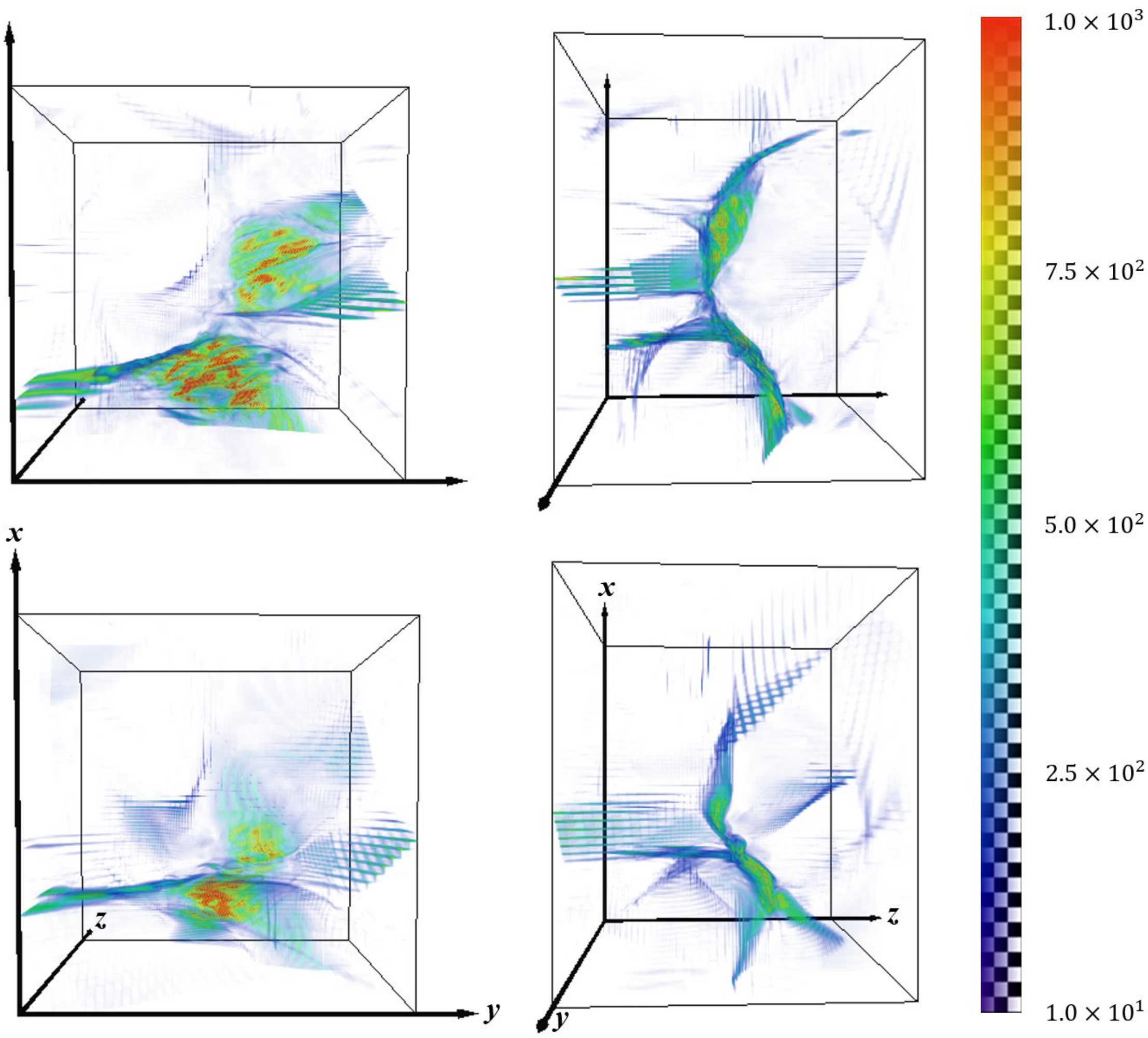}
\caption{3D images of the current density magnitude normalized by $\sqrt{P_0}/L$  at $t=1.3L/C_s$ (upper row) and $t=2.0L/C_s$ (lower row) with two different viewing angles of the ABC field. The domain of these images is $[L/4,5L/8] \times [0,3L/8] \times [0,5L/16]$ in the rest frame . Current sheets are produced shortly before $t=1.3L/C_s$ and two adjacent current sheets are in the process of merging at $t=2.0L/C_s$.  The strongest current densities are distributed over several patches on thin sheets.}
\label{fig: ABC_flow_3D_image}
\end{figure*}
\begin{figure*}
\centering
\includegraphics[scale=0.54]{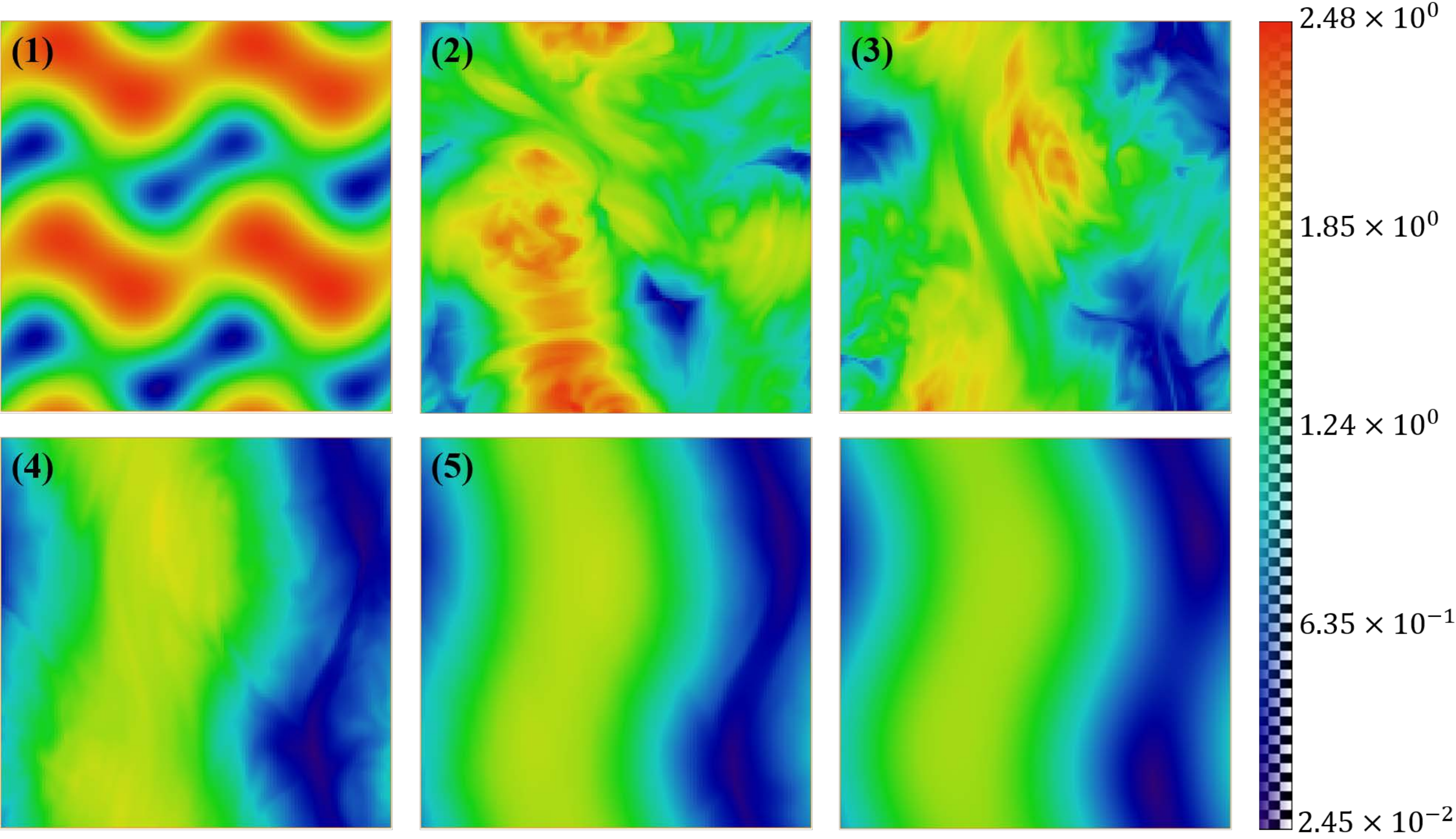}
\caption{Time sequence of magnetic field magnitude slice images of the ABC field. The slice is cut at $y=0.77L$ in the rest frame. Numbers labeled in the figure are in the chronological order at $t=0$, $3.9$, $7.7$, $15.5$ and $31.0$, respectively. Here time is normalized by $L/C_s$. The color stands for the magnetic field magnitude. The right bottom image is constructed using Eq. (\ref{equ: fitting final stage formula}), the one-period ABC configuration. The reconstructed image agrees with the $5th$ image very well, and the system converges to a stable state.}
\label{fig: Taylor_conjecture}
\end{figure*}
\begin{figure}
\includegraphics[scale=0.355]{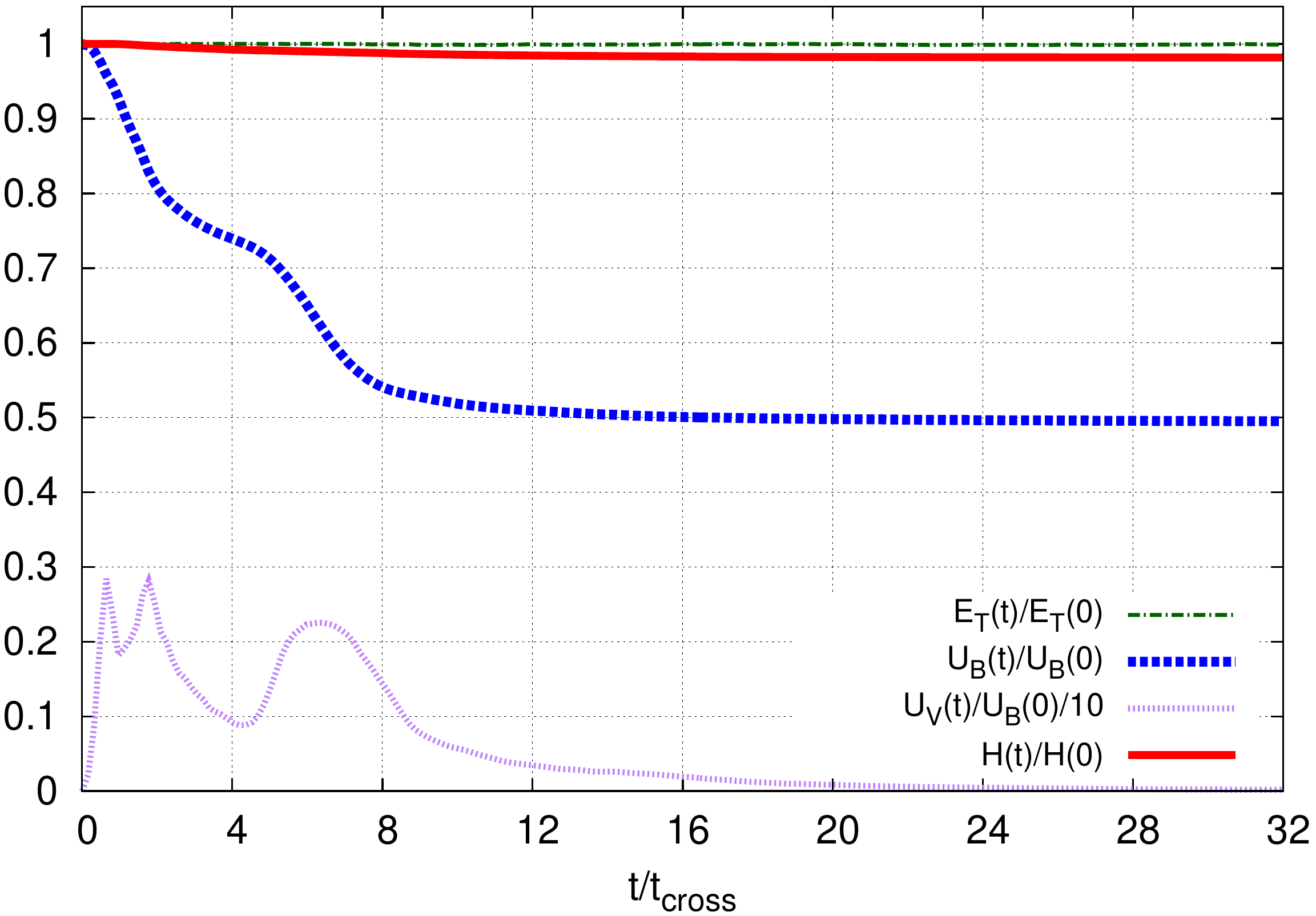}
\caption{Time evolution of the total energy $E_T$ (dash-dotted line), the magnetic energy $U_B$ (dashed line), the flow energy $U_V$ (dotted line)  and the magnetic helicity $H$ (solid line) of the  ABC field. The horizontal axis is the time normalized to the sound crossing time ($t_{cross} = L/C_s$). The total, magnetic energy and helicity are normalized to their corresponding initial values while the flow energy is normalized to $U_B(0)/10$ with $U_B(0)$ being the initial magnetic energy for comparison. The total energy remains the same due to the finite-volume method. The magnetic energy decreases to a half of $U_B(0)$, and the flow energy initially increases and then decays. The magnetic helicity remains approximately constant ($ > 98\%$) up to $32$ sound crossing time.}
\label{fig: magnetic_energy_helicity}
\end{figure}

This is a new test problem demonstrating 3D relaxation for a MHD system.  Due to its intrinsic periodic boundary condition, this problem is also ideal for examining the weak scaling performance for a parallelized code. Inspired by the intriguing Arnold-Beltrami-Childress (ABC) flow, which is a three-dimensional incompressible flow \citep*{Arnold1965, Childress1970}, we examine turbulent relaxation of magnetic field with the ABC configuration in a uniform plasma with density $\rho_0$ and pressure $P_0$. The magnetic field pattern is $\bm{B}_0(=(B_{x,0},B_{y,0},B_{z,0}))$ and
\begin{equation}
\label{equ: B_0 in performance test}
\left.\begin{aligned}
& B_{x,0} = A\sin\Big ( {{2\pi n}\over L} z \Big ) + C\cos \Big ( {{2\pi n}\over L} y \Big ),\\
& B_{y,0} = B\sin\Big ( {{2\pi n}\over L} x \Big ) + A\cos \Big ( {{2\pi n}\over L} z \Big ),\\
& B_{z,0} = C\sin\Big ( {{2\pi n}\over L}y \Big  ) + B\cos \Big ( {{2\pi n}\over L} x \Big ),
\end{aligned}
\right.
\end{equation}
where the cubic box length $L$ is set to unity, $A$, $B$ and $C$ are constants and $n$ is an integer governing the number of periods in the box.  It is noted that this configuration is a force-free equilibrium since the current density $\bm{J}_0 \equiv \nabla\times\bm{B_0} = 2\pi n\bm{B_0}/L$ is parallel to the magnetic field.  As shown in Appendix, one finds the configuration with $n=1$ is stable.  Therefore, we choose $n=2$ state as the initial equilibrium.  In addition, we choose $A=B=C=\sqrt{P_0}$, i.e., plasma $\beta=1$.

We also let this plasma move with a small uniform drift velocity along the diagonal direction, i.e., the drift velocity $\bm{V}_0 = (V_{x,0},V_{y,0},V_{z,0})$ with $V_{x,0}=V_{y,0}=V_{z,0}$ and $V_{x,0} > 0$. We choose the drift velocity $8\sqrt{\gamma}$ times smaller than the sound speed $C_s \equiv \sqrt{\gamma P_0/\rho_0}$.  In principle, this uniform flow doesn't affect the stability due to Galilean invariance of MHD equations. However, the static grids can weakly break the Galilean invariance, depending on numerical schemes.  When the plasma is static, there is an instability of long and thin (grid-scale) structures produced in the simulation.  This instability still persists in the $n=1$ equilibrium state, which is proven to be ideal MHD stable, as shown in Appendix.  Similar unstable structures are also produced in the ATHENA code. The physics underlying these long and thin structures is interesting, and it may originate from resistive instability despite no resistivity is explicitly used in these codes.   It is well known that fast resistive tearing mode instabilities can occur in an ideal MHD marginally stable plasma, e.g. \citet{Rosenbluth1973}.  We shall further investigate on this instability in a future work.  For now, we find the uniform flow can numerically suppress these small-scale instabilities and yields only physical large-scale ideal MHD instabilities.

Finally, small but finite-amplitude initial perturbations are added on the equilibrium configuration to speed up the uninteresting linear phase for high-resolution runs. In order to find appropriate perturbations, we start with the above initial configuration with no explicitly added noise in a low resolution simulation.  An unstable $n=1$ perturbation then arises out of numerical noise.  We analyze such a unstable magnetic field perturbation in the linearly regime and identify the largest four Fourier components of the unstable $\delta\bm{B}$, which is supposed to be the most unstable eigenmode:
\begin{equation}
\label{equ: perturbed Bx}
\left.\begin{aligned}
{{\delta B_x}\over{\sqrt{P_0}}} = 10^{-2}\Big \{  & + 2.37\cos\Big [ {{2\pi}\over L} (x+y) \Big ] + 2.10\sin\Big [ {{2\pi}\over L} (x+y) \Big ] \\
                                          & - 3.48\cos\Big [ {{2\pi}\over L} (x+z) \Big ] - 3.18\sin\Big [ {{2\pi}\over L} (x+z) \Big ] \\
                                          & - 5.86\cos\Big [ {{2\pi}\over L} (z-y) \Big ] + 5.60\sin\Big [ {{2\pi}\over L} (z-y) \Big ] \\
                                          & + 2.08\cos\Big [ {{2\pi}\over L} (x-z) \Big ] + 1.94\sin\Big [ {{2\pi}\over L} (x-z) \Big ]  \Big \},
\end{aligned}
\right.
\end{equation}
\begin{equation}
\label{equ: perturbed By}
\left.\begin{aligned}
{{\delta B_y}\over{\sqrt{P_0}}} = 10^{-2}\Big \{  & - 2.37\cos\Big [ {{2\pi}\over L} (x+y) \Big ] - 2.10\sin\Big [ {{2\pi}\over L} (x+y) \Big ] \\
                                          & - 4.90\cos\Big [ {{2\pi}\over L} (x+z) \Big ] + 5.37\sin\Big [ {{2\pi}\over L} (x+z) \Big ] \\
                                          & + 3.60\cos\Big [ {{2\pi}\over L} (z-y) \Big ] + 3.80\sin\Big [ {{2\pi}\over L} (z-y) \Big ] \\
                                          & - 3.00\cos\Big [ {{2\pi}\over L} (x-z) \Big ] + 3.22\sin\Big [ {{2\pi}\over L} (x-z) \Big ] \Big \},
\end{aligned}
\right.
\end{equation}
\begin{equation}
\label{equ: perturbed Bz}
\left.\begin{aligned}
{{\delta B_z}\over{\sqrt{P_0}}} = 10^{-2}\Big \{ & - 3.23\cos\Big [ {{2\pi}\over L} (x+y) \Big ] + 3.66\sin\Big [ {{2\pi}\over L} (x+y) \Big ] \\
                                         & + 3.48\cos\Big [ {{2\pi}\over L} (x+z) \Big ] + 3.18\sin\Big [ {{2\pi}\over L} (x+z) \Big ] \\
                                         & + 3.60\cos\Big [ {{2\pi}\over L} (z-y) \Big ] + 3.80\sin\Big [ {{2\pi}\over L} (z-y) \Big ] \\
                                         & + 2.08\cos\Big [ {{2\pi}\over L} (x-z) \Big ] + 1.94\sin\Big [ {{2\pi}\over L} (x-z) \Big ] \Big \}.
\end{aligned}
\right.
\end{equation}
Note that Eqs. (\ref{equ: perturbed Bx}), (\ref{equ: perturbed By}) and (\ref{equ: perturbed Bz}) yields a perturbed magnetic field also satisfying the divergence-free constraint, and that the wave numbers of the perturbation eigenmode rotate $\pm  45$ degrees from the equilibrium field with $\sqrt{2}$ longer wavelength. We then add the magnetic field perturbation into the $n=2$ force-free equilibrium as the initial condition for high-resolution runs.

The numerical schemes are the same as the previous blast wave problem (except for adopting the single-precision calculation), and the periodic boundary condition is used. The quantity $|\bm{J}|/|\bm{B}|$ is adopted as the refinement criterion and the grid with this quantity higher than $24/\Delta h(k)$ is refined to level $k+1$, where $k=1$, $2$, $3$ and $4$, and $\Delta h(k)$ is the $k$-th level grid size.  Similar to the Orszag-Tang vortex test, this criterion aims to capture magnetic reconnection.

Figure (\ref{fig: ABC_flow_3D_image}) is the 3-D image of the current density magnitude $|\bm{J}|$ (normalized by $\sqrt{P_0}/L$) at $t=1.3L/C_s$ (upper row) and $t=2.0L/C_s$ (lower row) with two different orthogonal orientations. This image is obtained by an AMR simulation that has $128^3$ grids at the base level with $4$ levels of refinement. We find the total number of grid cells and the volume-filling fraction on each refinement level decrease monotonically from $t=1.3L/C_s$ to $t=2.0L/C_s$. For example, there are about $3.5 \times 10^7$ grid cells and the volume-filling fractions from the lowest to the highest refinement levels are $18.7\%$, $3.6\%$, $0.9\%$ and $0.2\%$, respectively, at $t=1.3L/C_s$. However, when $t=2.0L/C_s$, these numbers drop to $2.4 \times 10^7$, $11.0\%$, $2.2\%$, $0.5\%$ and $0.1\%$, respectively.

Figure (\ref{fig: ABC_flow_3D_image}) reveals strong current densities occurring in very thin sheets, and the two adjacent current sheets merge together in less than one sound crossing time. The magnetic energy is found to decrease dramatically during the merging process (c.f. Fig. (\ref{fig: magnetic_energy_helicity})). Therefore, magnetic reconnection must occur at the appearance of strong current sheets. It is however difficult to quantify the local reconnection rate of such 3D reconnection, unlike Sweet-Parker or Petschek reconnection \citep*{Petschek1964} in 2D, since the strongest current densities appear in several relatively short-lived patches on the sheet.  The patch reconnection occurs not at the magnetic nulls but in weak field regions.  This result is at variant with the 3D separator reconnection proposed before \citep*{Parnell2010}, which is filament-like reconnection.  It remains to be seen whether the patchy sheet reconnection is generic in 3D MHD.

We also examine the long-term evolution of this unstable force-free equilibrium and find the system relaxes to another equilibrium. Figure (\ref{fig: Taylor_conjecture}) shows the time evolution of a slice image of the magnetic field magnitude.  Over $31$ sound crossing time ($31L/C_s$), the system approaches an $n=1$ equilibrium state which can be fitted by the following formula
\begin{equation}
\label{equ: fitting final stage formula}
\left.\begin{aligned}
& {B_{x}\over{\sqrt{P_0}}} = 0.33\sin \Big [ {{2\pi}\over L}(z-0.68) \Big ] + 0.80\cos \Big [ {{2\pi}\over L}(y-0.50) \Big ],\\
& {B_{y}\over{\sqrt{P_0}}} = 0.85\sin \Big [ {{2\pi}\over L}(x-0.36) \Big ] + 0.33\cos \Big [ {{2\pi}\over L}(z-0.68) \Big ],\\
& {B_{z}\over{\sqrt{P_0}}} = 0.80\sin \Big [ {{2\pi}\over L}(y-0.50) \Big ] + 0.85\cos \Big [ {{2\pi}\over L}(x-0.36) \Big ].
\end{aligned}
\right.
\end{equation}
Eq. (\ref{equ: fitting final stage formula}) shows that the relaxed state is again an ABC configuration with only one period.

The respective evolutions of total energy ($E_T \equiv \int e d^3\bm{x}$ with the total energy density $e$ defined in Sec. (\ref{sec:MHD Equations})), magnetic energy ($U_B \equiv \int |\bm{B}|^2/2d^3\bm{x}$), flow energy ($U_V\equiv \int \rho|\bm{V}|^2/2 d^3\bm{x}$) and magnetic helicity ($H \equiv \int \bm{A}\cdot\bm{B}d^3\bm{x}$ with the vector potential $\bm{A}$) are shown in Fig. (\ref{fig: magnetic_energy_helicity}). As expected, the total energy remains constant to the machine precision.  By contrast, the magnetic energy is monotonically decreasing, while the flow energy first increases dramatically and then oscillates. However, the flow energy is dissipated to zero in the end. On the other hand, the magnetic helicity, which is also a conserved quantity in ideal MHD and related to the linkage of magnetic field lines, only drops by about $2\%$ up to $32$ sound crossing times.  Moreover, the magnetic energy for the final relaxed state is about twice smaller than the initial state.  This can be derived from the conservation of the magnetic helicity, as follows.

Note that the initial and final states follow the ABC patterns. Straightforward calculations yield the magnetic energy and the magnetic helicity with the ABC field pattern to be $U_B = (A^2+B^2+C^2)L^3/2$ and $H = (A^2+B^2+C^2)L^4/(2\pi n) = U_BL/(\pi n)$, respectively.  Since the magnetic helicity is a constant, we conclude $U_B \propto n$, and hence the magnetic energy for the final state should be twice smaller than the initial state as $n$ changes from $2$ to $1$. The above analysis confirms the Taylor's conjecture that unstable MHD systems tend to minimize magnetic energy subject to the constraint of a constant magnetic helicity \citep*{Taylor1986}. The minimum-energy state is a force-free state with a uniform current-to-field ratio, and our final ABC configuration is such a minimum-energy state.

\subsection{performance test}
\label{subsec: performance test}

In this section, we demonstrate the performance of GAMER-MHD, which will be divided into two parts. First, we measure the performance of GPU (and CPU) MHD solver alone without any AMR operation. Second, we measure the overall performance of strong and weak scaling tests including AMR. The initial condition, boundary condition, and AMR setups all follow (\rmnum{5}) of Sec. (\ref{subsec: accuracy test}).

\bigskip

(\rmnum{1}) GPU performance

\medskip

\begin{figure}
\includegraphics[scale=0.35]{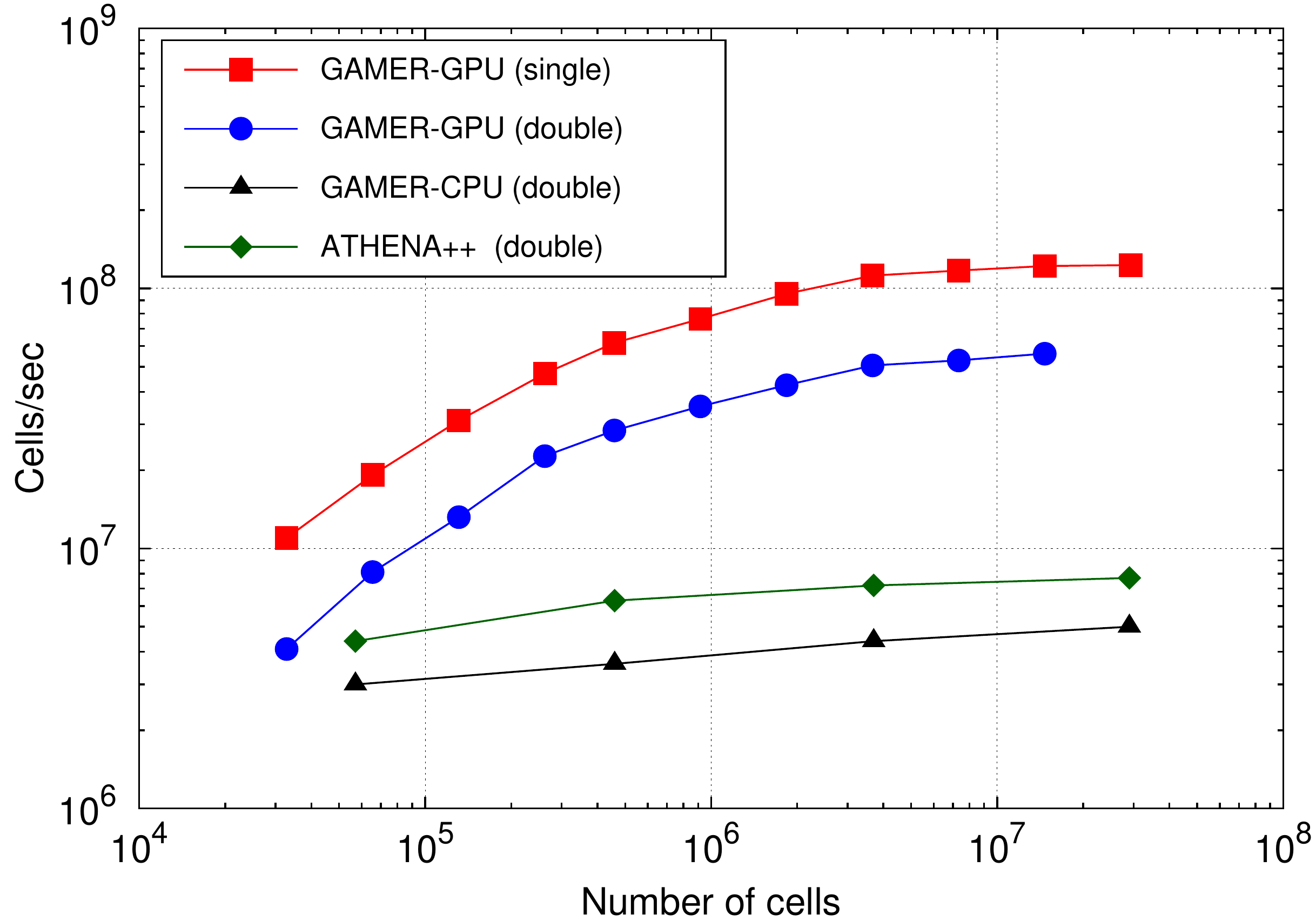}
\caption{Performances of a single GPU and CPU. We measure the GPU performance on a single Tesla P100 GPU with both single precision (squares) and double precision (circles). Also plotted are the CPU performances of GAMER-CPU (triangles) and ATHENA++ (diamonds) with double precision conducted on an Intel Xeon E5-2680 14-core processor (using all 14 cores) using the same MHD scheme.}
\label{fig: GPU performance}
\end{figure}

We adopt the number of cell-updates per second for quantifying the performance and measure it by simulating different resolutions of uniform grids on a single Tesla P100 GPU. In order for directly comparing with the ATHENA++ code \citep{Stone2008, White2016}, we adopt the same VL scheme with PLM reconstruction and Roe's solver as those implemented in ATHENA++.

Figure (\ref{fig: GPU performance}) shows the single-GPU performance.  We measure the performance of both single- and double-precision calculations. The performance is a monotonic function of the total cell numbers to be updated and saturates when the cell number is sufficiently large. The GPU performance starts to saturate when the total cell number exceeds $\sim 2 \times 10^6$ and reaches$\sim 1.2 \times 10^8$ and $5.5 \times 10^7$ cell-updates per second for single- and double-precision calculations, respectively. The single-precision performance is about $2$ times faster than the double-precision performance on a Tesla P100 GPU. We also measure the CPU performance on an Intel Xeon E5-2680 14-core processor with the double-precision calculation using the same MHD scheme.  The CPU performance is saturated at about $5.0 \times 10^6$ cell-updates per second using all 14 cores. Finally, we also measure the performance of ATHENA++ and it reaches $7.7 \times 10^6$ cell-updates per second, $1.5$ times faster than our CPU code, likely due to the more optimized implementation of vectorization\footnote{On an Intel Knights Landing (KNL) machine, ATHENA++ achieves $5 \times 10^7$ and $3 \times 10^7$ cells/sec for single and double precision, respectively (James Stone, private communication), about half of the GAMER-MHD performance measured on a P100 GPU.}. Further optimizations for the CPU performance of GAMER will be investigated in the near future.

\bigskip

(\rmnum{2}) Overall performance

\medskip

\begin{figure}
\includegraphics[scale=0.355]{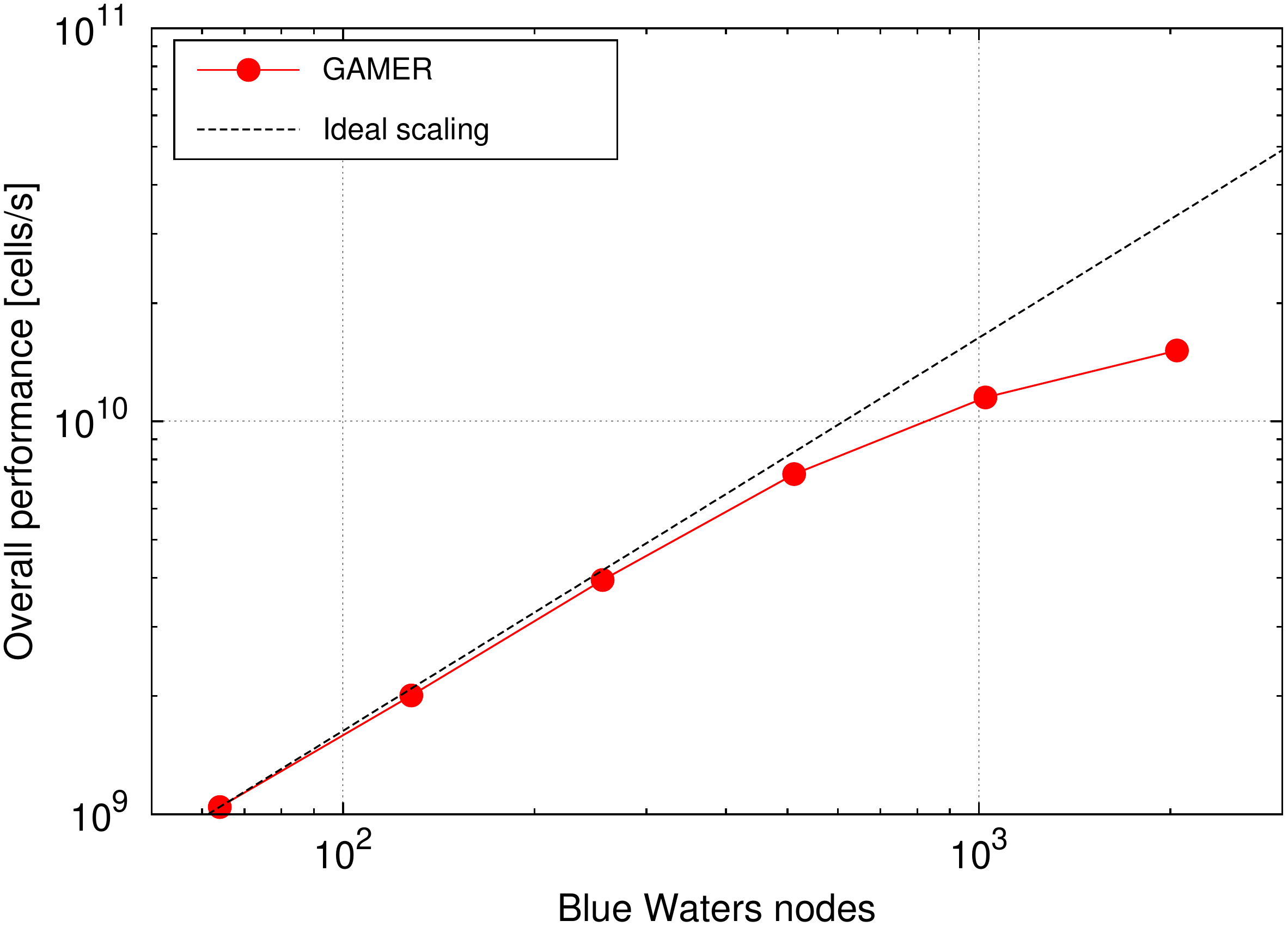}
\caption{Overall performance for the strong scaling test. The horizontal axis is the number of computing nodes and the vertical axis is the number of cell-updates per second. Filled circle points are the measured performance and the dashed line is the ideal scaling, i.e., single GPU (K20X) performance $\times$ node number. The performance closely follows the ideal scaling law when using less than 1,024 nodes. More quantitative analyses are in Fig. (\ref{fig: Strong scaling}).}
\label{fig: strong_scaling_overall_performance}
\end{figure}
\begin{figure*}
\centering
\includegraphics[scale=0.73]{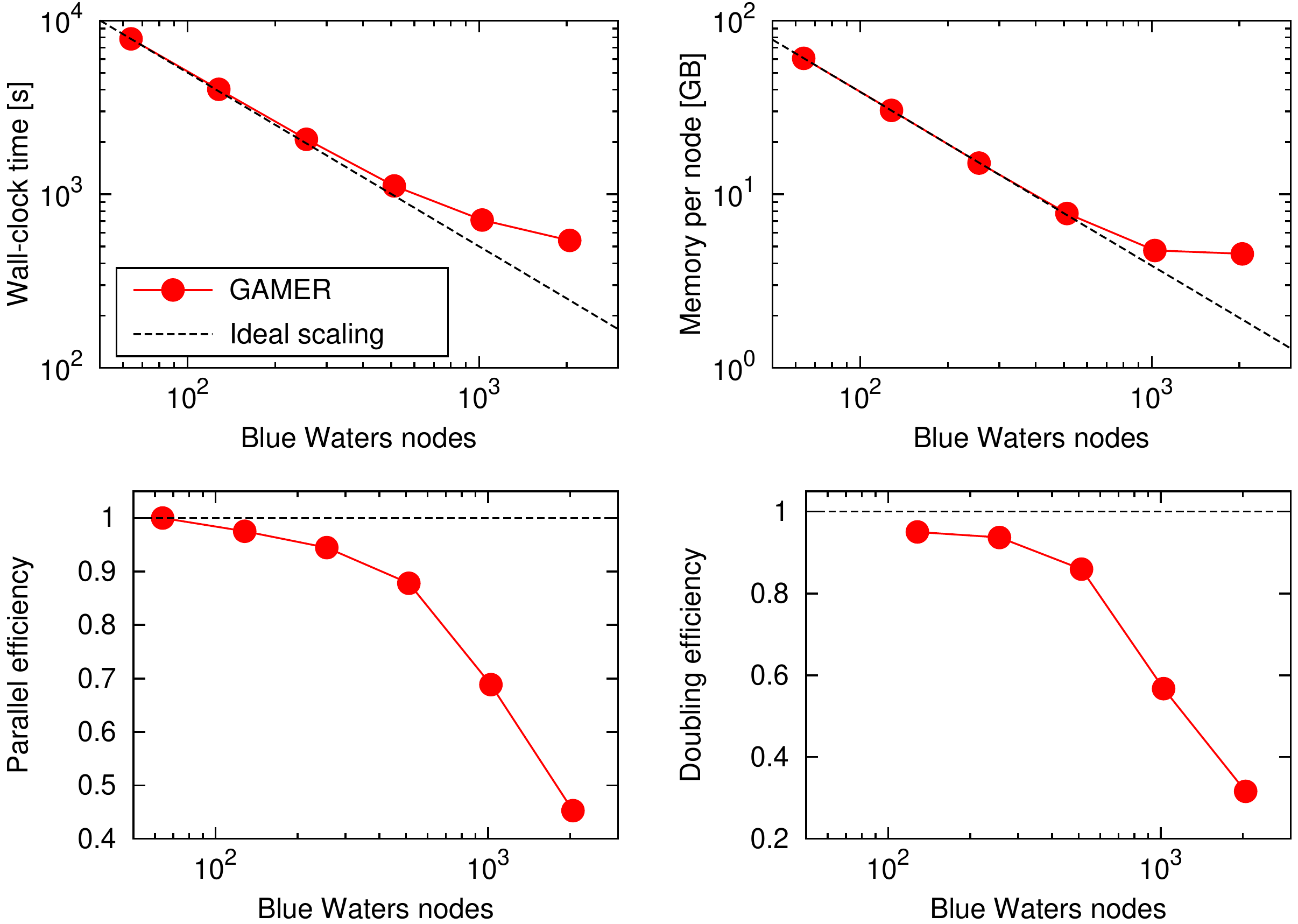}
\caption{Quantitative analysis of the strong scaling of GAMER-MHD. The horizontal axis is the number of Blue water computing nodes. Four panels are (1) wall-clock time top left, (2) memory per node top right, (3) parallel efficiency bottom left, and (4) doubling efficiency bottom right. Definitions of these quantities are in the text. The code achieves $70\%$ parallel efficiency and $57\%$ doubling efficiency with $10^3$ computing nodes.}
\label{fig: Strong scaling}
\end{figure*}
\begin{figure}
\includegraphics[scale=0.355]{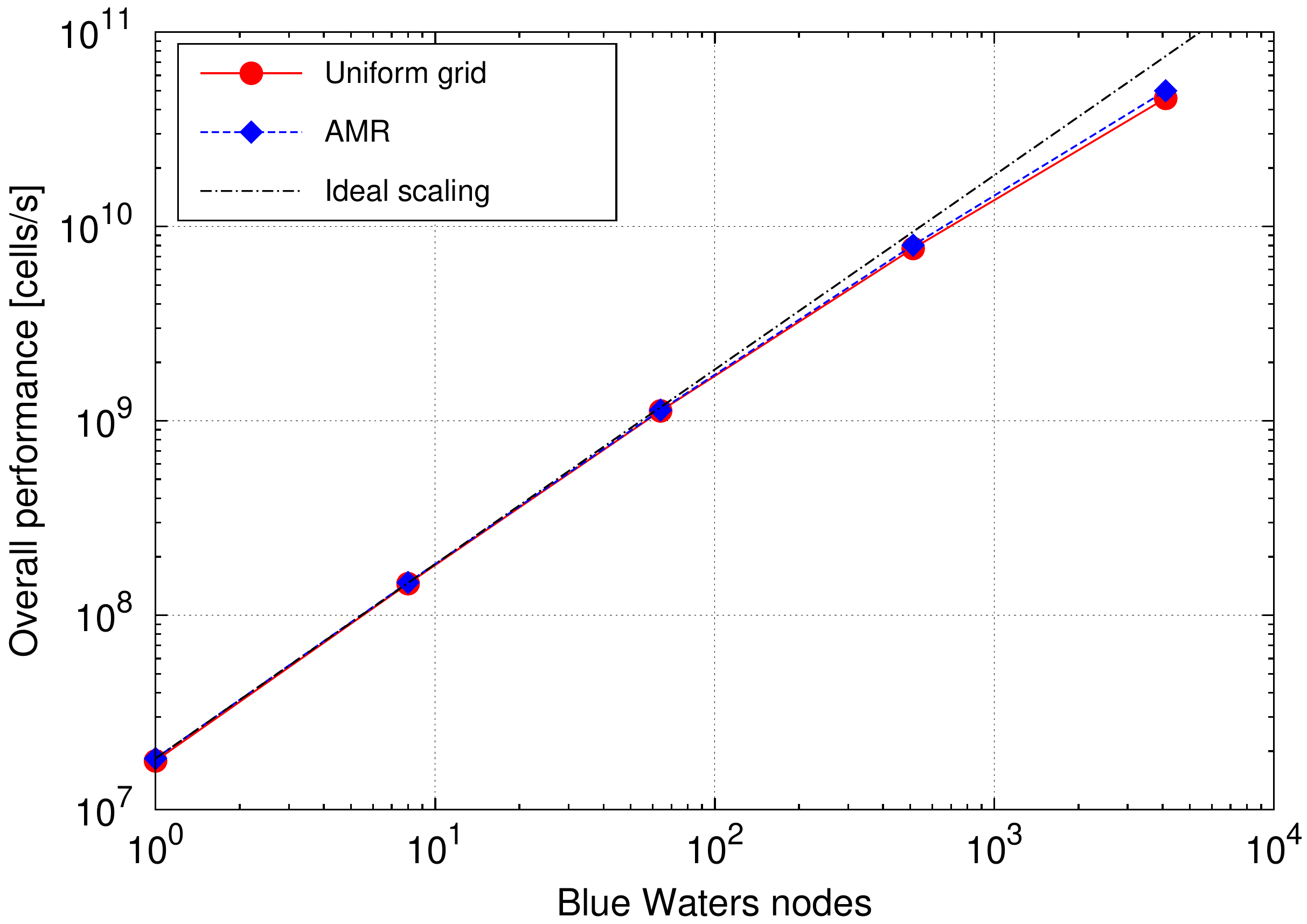}
\caption{Overall performance for the weak scaling test. The horizontal and vertical axes are the same with Fig. (\ref{fig: strong_scaling_overall_performance}). Filled circles are the uniform-grid performance and filled diamonds are the AMR performance. The dashed-dotted line is the ideal scaling. The AMR performance is almost the same as the uniform-grid case and closely follows the ideal scaling law up to $4096$ computing nodes. More quantitative analysis is in Fig. (\ref{fig: Weak scaling}).}
\label{fig: weak_scaling_overall_performance}
\end{figure}
\begin{figure*}
\centering
\includegraphics[scale=0.73]{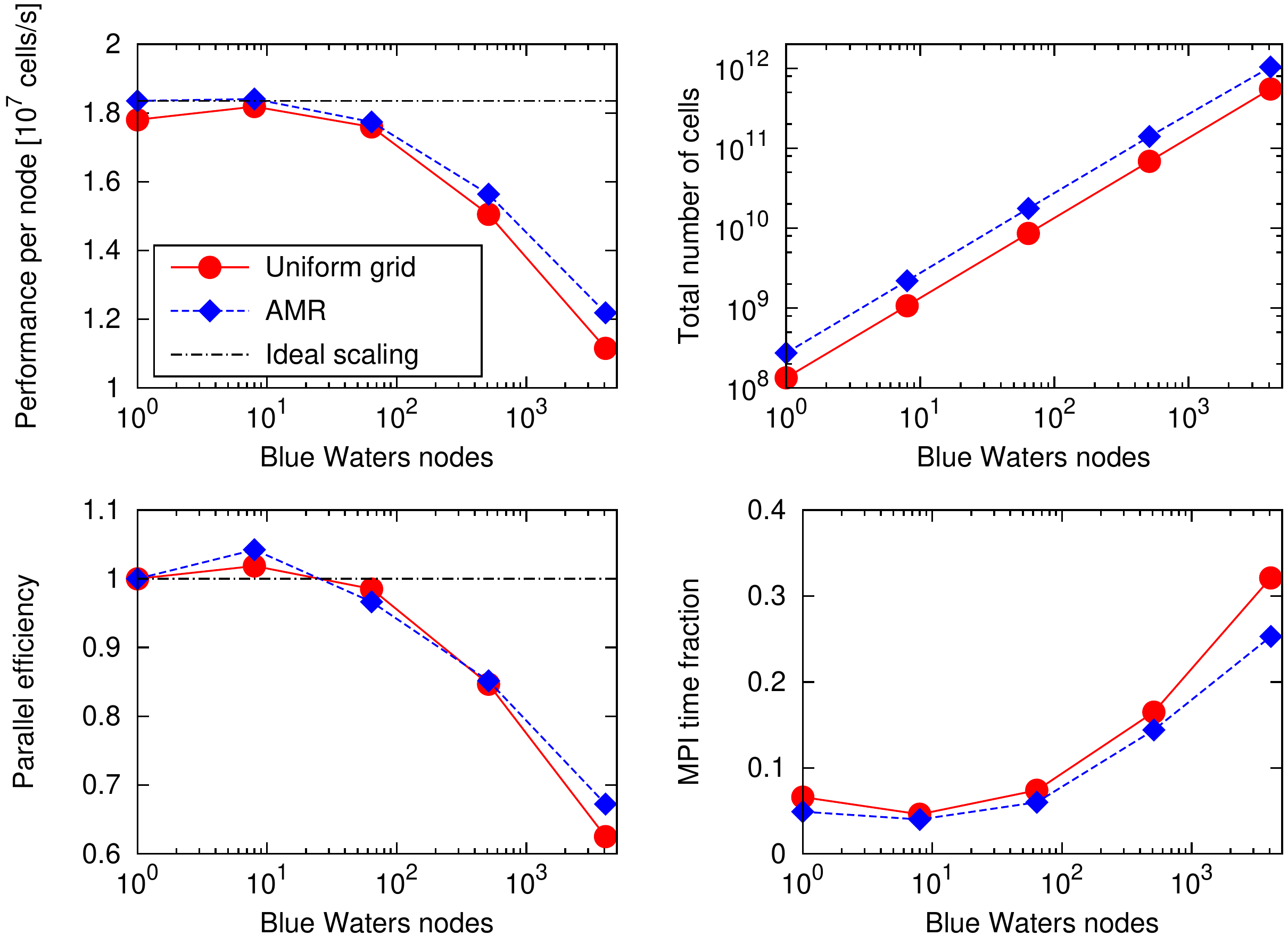}
\caption{Quantitative analysis of the weak scaling. Four panels are (1) number of cells updated per second per node top left, (2) total number of cells top right, (3) parallel efficiency bottom left, and (4) MPI time fraction bottom right. The code achieves $85\%$ parallel efficiency with $512$ computing nodes.}
\label{fig: Weak scaling}
\end{figure*}

In this performance test we measure the overall efficiency of GAMER-MHD with AMR and hybrid MPI/OpenMP/GPU parallelization. We conduct both strong and weak scaling on the XK nodes of the Blue Waters supercomputer, where each computing node is equipped with an AMD Opteron 6276 CPU with 16 processing cores and a Tesla K20X GPU. Here we turn to the CTU scheme with PPM reconstruction, Roe's solver and single-precision calculation to measure the scaling behavior.

The total number of cells is fixed in the strong scaling test, and thus, ideally, the total simulation time should be inversely proportional to the number of computing nodes. However, it is impossible to achieve ideal performance with a very larger number of computing nodes due to the insufficient number of cells in each GPU (see Fig. (\ref{fig: GPU performance})).  Moreover, the execution time of AMR operations (see Sec. (\ref{subsec:Adaptive Mesh Refinement})) and MPI communications will become non-negligible for a large node number. All these will decrease the overall performance. Therefore, the strong scaling is to find the sweet spot of the computing node number that utilizes the computational resource efficiently.  Figure (\ref{fig: strong_scaling_overall_performance}) shows the strong scaling overall performance. The overall performance scales reasonably well from $64$ to $1,024$ nodes and still reaches $1.5 \times 10^{10}$ cell-updates per second for $2048$ nodes.

We stress that the average performance per node when using 64 nodes is only about $0.82$ times lower than the GPU performance. The AMR operations (c.f. Sec. (\ref{subsec:Adaptive Mesh Refinement})) and MPI communication are measured to take about $9\%$ and $6\%$ of the total time, respectively, indicating that the deviation from the ideal single-GPU performance is caused by AMR operations and MPI communication. This result also demonstrates a very efficient overlap between the GPU and CPU concurrent executions (c.f. Sec. (\ref{sec: GAMER MHD Structure})).

Figure (\ref{fig: Strong scaling}) shows the detailed quantitative analysis for the strong scaling test. Two quantities are introduced in this figure, parallel efficiency and doubling efficiency. The former marks the efficiency of $M$ nodes compared with the minimum node number $M_{min}$, and the latter indicates the performance gain by increasing the node number by a factor of $2$. While the former is the conventional measure of performance of a large node number, the latter is a new measure and more practical for users to determine the optimal node number for a given problem size as this is an indicator near the saturation end. Specifically, the parallel efficiency is defined as $[T(M_{min})/T(M)]/(M/M_{min})$, in which $T(M)$ is the wall-clock time with $M$ nodes when a calculation is executed for a fixed amount time steps, and the doubling efficiency is defined as $T(M/2)/T(M)-1$. This definition of doubling efficiency approaches zero when there is no gain in using $M$ nodes compared with using $M/2$ nodes.

Figure (\ref{fig: Strong scaling}) shows $70\%$ parallel efficiency and $57\%$ doubling efficiency with $10^3$ computing nodes. The wall-clock time and the memory per node deviate from the ideal scaling by $14\%$ and $2\%$ for $512$  nodes and by $45\%$ and $25\%$ for $1024$ nodes, respectively. We also found the MPI communicating time for $2048$ nodes is only about $5$ times smaller than $64$ nodes and the refinement/derefinement operations in AMR (c.f. Sec. (\ref{subsec:Adaptive Mesh Refinement})) for $2048$ nodes is about $6$ times smaller than $64$ nodes. The above two operations deviated from the scaling law, and are the main cause of the parallel efficiency drop.

Following the replicated problem technique proposed by \citet{Calhoun2017}, the ideal case for weak scaling has the problem size, and thus the total number of cells, proportional to the number of nodes. Therefore, unlike the strong scaling test, the wall-clock time should ideally be independent of the number of nodes.  Figure (\ref{fig: weak_scaling_overall_performance}) shows the overall performance for the weak scaling test. The performance scales reasonably well from one to thousands of nodes and achieves $5 \times 10^{10}$ cell updates per second for $4096$ nodes. The overall performance with AMR is almost the same as the uniform-grid simulation, revealing that the extra works in the AMR application take only an insignificantly small execution fraction of time. For reference, when running with $4,096$ nodes, the uniform-grid simulation has a resolution of $8,192^3$ cells and the AMR case adopts a base-level resolution of $2048^3$ cells with $4$ additional refinement levels.

Figure (\ref{fig: Weak scaling}) reveals the detailed quantitative analysis for the weak scaling test. Here the parallel efficiency is defined as $T(M_{min})/T(M)$. It still achieves $70\%$ parallel efficiency when updating $2 \times 10^{11}$ cells with $10^3$ nodes. Note that the MPI time fraction increases noticeably when $M \geq 512$, which is the main cause of the drop of parallel efficiency. One plausible reason for the increasing MPI time fraction is the network topology of Blue Waters.  This fraction may be decreased by overlapping the MPI communication with both CPU (AMR operations) and GPU (MHD solver) computations, which will be investigated in the future.

\section{conclusion}
\label{sec: conclusion}

In this work, we demonstrate successful implementation of the MHD scheme (CTU+CT and VL+CT) into GAMER. The Balsara's method is also implemented in AMR to maintain the divergence-free constraint of the magnetic field. By taking advantage of the GAMER infrastructure, the MHD scheme can be accelerated by GPU and the AMR operations are executed in parallel by multi-core CPU via OpenMP.  The data transfer between CPU and GPU memory is overlapped by the execution of the GPU MHD scheme via {\it{CUDA stream}}.  The preparation and closing steps performed in the multi-core CPU is concurrent with the GPU execution. Finally, GAMER-MHD can run in parallel via MPI, where the Hilbert space-filling curve decomposition is used for load balancing.

Our performance test results show the following. The single Tesla P100 GPU performance of the MHD solver can reach $1.2 \times 10^8$ cell updates/sec for the single-precision calculation and $5.5\times 10^7$ cell updates/sec for the double-precision calculation. This performance is achieved even in complicated AMR applications due to the efficient overlapping between the CPU and GPU operations. Furthermore, we demonstrate a parallel efficiency of $ \sim 70\%$ for both weak and strong scaling using $1,024$ nodes on the Blue Waters supercomputer.   

It is possible to optimize the GAMER-MHD performance further. For example, the MPI communication time may be reduced by taking advantage of the non-blocking communication. Also, the CPU performance may be improved further by explicitly leveraging vectorization. These optimizations will be investigated in the near future.

We have adopted the linear wave and shock-tube tests to verify the correctness of the MHD scheme implementation.  The Orszag-Tang vortex and the blast wave aim to test AMR implementation, and we find the results can reproduce those with high-resolution uniform-grid simulations. Taking advantage of the AMR strategy, very high resolution simulations are easily achievable for these test problems with reasonable run times, in which interesting small-scale features, undetected before, are found. For example, in the Torrilhon shock tube test the erroneous compound wave appearing in low resolution simulations can be eliminated by increasing the resolution, and the numerical solution indeed converges to the exact solution albeit at a slow rate; magnetic reconnection in the two-dimensional Orszag-Tang Vortex test are identified to be a variant of Sweet-Parker type; in the blast-wave test, weak three-dimensional small-scale instabilities are discovered near the slow shock front parallel to the magnetic field.

We have explored a new test problem, a fully three-dimensional MHD equilibrium of the Arnold-Beltrami-Childress (ABC) force-free magnetic field configuration, for testing the scaling of overall computation performance.  This study yields a physics result consistent with the Taylor's conjecture that higher-energy MHD equilibria subject to instabilities can relax to a unique lowest-energy force-free state preserving the magnetic helicity.   This unique configuration is attainable after a turbulent phase, through which tangled magnetic field lines are broken via reconnection, and the system finally reaches a configuration with large-scale current-carrying force-free field, rather than a current-free potential field.  This minimum energy state still carries current due to a global field line linkage constraint, i.e., a constant global magnetic helicity, a topological constraint that limits the minimum energy which the system can assume. We also find a new 3D magnetic ”patch” reconnection during turbulent relaxation, a variant of sheet connection that is very different from the thin-tube reconnection associated with the magnetic separators \citep*{Parnell2010}.
Whether the sheet structure is generic in 3D magnetic reconnection remains to be investigated.

\section*{Acknowledgements}
H. S. is grateful to James Stone for insightful discussions. This project is supported in part by MOST of Taiwan under the grant MOST 103-2112-M-002-020-MY3. It is also part of the Blue Waters sustained-petascale computing project, which is supported by the National Science Foundation (awards OCI-0725070 and ACI-1238993) and the state of Illinois. Blue Waters is a joint effort of the University of Illinois at Urbana-Champaign and its National Center for Supercomputing Applications (NCSA). This work also used computational resources provided by the Innovative Systems Lab (ISL) at NCSA.

This publication is also supported in part by the Gordon and Betty Moore Foundation's Data-Driven Discovery Initiative through Grant GBMF4561 to Matthew Turk, and is based upon work supported by the National Science Foundation under Grant No. ACI-1535651.

We acknowledge the following codes that are relevant to this work: \software{ GAMER (\citealt{bSc1, bSc2, Schive2017}; \url{https://github.com/gamer-project/gamer}), ATHENA (\citealt{Stone2008}), ATHENA++ (\citealt{White2016}; \url{https://github.com/PrincetonUniversity/athena-public-version}).}

\renewcommand{\theequation}{A\arabic{equation}}
\setcounter{equation}{0}
\section*{appendix: Stability of ABC magnetic field}
\label{sec:appendix}
Here we have the equilibrium magnetic field defined in Eq. (\ref{equ: B_0 in performance test}), where $\bm{J}_0 = 2\pi n\bm{B_0}/L$. For simplicity, we let $L=1$.

The linear stability is revealed by applying time-domain Fourier transformation on linearized MHD equation, yielding the eigenvalue problem:
\begin{equation}
\label{equ: displacement equation}
-\rho_0 \omega^2 \bm{\xi} = \gamma P_0 \nabla^2\bm{\xi} + \alpha_n \bm{B}_0 \times [\nabla \times (\bm{\xi} \times \bm{B}_0)] + \{\nabla \times [\nabla \times (\bm{\xi} \times \bm{B}_0)]\} \times  \bm{B}_0,
\end{equation}
where $\bm{\xi}$ is the displacement defined as $-i\omega \bm{\xi} \equiv \bm{\delta v}$ with the perturbed velocity field $\bm{\delta v}$, and $\alpha_n \equiv 2\pi n$.

Note that the right hand side of Eq. (\ref{equ: displacement equation}) is a Hermitian operator respect to the eigenfunction $\bm{\xi}$. Therefore, eigenvalues ($\rho_0 \omega^2$) of Eq. (\ref{equ: displacement equation})  are real and the corresponding eigenfunctions are orthonormal. The system is stable if and only if all eigenvalues are positive.

We shall use the variational (energy) principle to prove the stability. Multiplying $\bm{\xi}^*$, the complex conjugate of $\bm{\xi}$, on both sides of Eq. (\ref{equ: displacement equation}) and then integrating both sides with the whole domain, we come up with
\begin{equation}
\label{equ: displacement equation integral form}
\rho_0 \omega^2 \int_{\Omega} |\bm{\xi}|^2 d^3\bm{x} = \gamma P_0 \int_{\Omega} |\nabla \cdot \bm{\xi}|^2 d^3\bm{x} + \int_{\Omega} |\nabla \times (\bm{\xi}\times \bm{B}_0)|^2 d^3\bm{x} - \alpha_n \int_{\Omega} (\bm{\xi}^{*}\times\bm{B}_0)\cdot[\nabla \times (\bm{\xi}\times \bm{B}_0)] d^3 \bm{x},
\end{equation}
where $\Omega$ is the whole domain.  Here the boundary condition is applied to eliminate boundary integral terms.  Hence the necessary and sufficient condition for stability becomes the right hand side of Eq. (\ref{equ: displacement equation integral form}) to be positive definite for any displacement $\bm{\xi}$.

In the following we shall first show that when the space is continuous and isotropic, a general result can be obtained.  We then proceed to the restricted situation where the space is a periodic cube, a realistic situation for simulations.

We now examine the stability of the $n=1$ state, i.e., $\alpha_n=2\pi$.  The first term on the right hand side of Eq. (\ref{equ: displacement equation integral form}) is the compressional term and positive definite, and therefore we only need to examine the remaining two terms.  We define $\bm{E} \equiv \bm{\xi} \times \bm{B}_0$, where $i\omega\bm{E}$ is the electric field.   When only the last two terms on the right of Eq. (\ref{equ: displacement equation integral form}) are considered, they can be regarded as functionals of $\bm{E}$.

The trivial marginally stable perturbation is $\bm{E}=0$ and the displacement is parallel to the field. More generally the mode has a displacement $\bm{\xi} \times \bm{B}_0 = \bm{C}$ for some constant vector $\bm{C}$.  When $\bm{C}=0$, it implies a field-aligned displacement $\bm{\xi}=f(\bm{x})\bm{B}_0$ for some scalar function $f(\bm{x})$.  Furthermore, if the displacement is incompressible, i.e., $\nabla\cdot\bm{\xi}=0$, it makes the compressional integral vanish and we can have a marginal stability. That is, the scalar function $f(\bm{x})$ is to satisfy $\bm{B}_0\cdot\nabla f=0$, meaning that the value of $f$ remains constant along any field line and different magnetic field lines may have different values of $f$.  This ground state equilibrium $\bm{B}_0$ has $8$ points where the magnetic field vanishes, and if $\bm{C}\neq 0$, $\bm{\xi}$ is singular at $\bm{B}_0 =0$, and the compressional integral diverges and thus it becomes vastly stable.

The non-trivial solution is the plane wave, which is the eigenfunction for $\bm{E}$.  Let $\bm{E_k}$ be the Fourier components $\bm{E}$, and the integrand becomes
\begin{equation}
\label{equ: continuous case Lorentz force term}
 |\bm{k} \times \bm{E_k}|^2+\alpha_n\Im[\bm{k}\cdot(\bm{E_k}^*\times\bm{E_k})] =(\bm{k}\times \bm{A})^2 +(\bm{k}\times\bm{D})^2 + 2\alpha_n\bm{k}\cdot(\bm{A}\times\bm{D}),
\end{equation}
where $\bm{A}=\Re[\bm{E_k}]$ and $\bm{D}=\Im[\bm{E_k}]$.  Now we let $\bm{A}$ and $\bm{D}$ lie on the $x-y$ plane and suspend an angle $\theta$ from each other.  That is, the first term is $k_z^2(A^2+D^2) + k_x^2(A_y^2+D_y^2)+k_y^2(A_x^2+D_x^2)$ and the second terms is $2\alpha_nk_z AD\sin\theta$, where $A\equiv |\bm{A}|$ and $D\equiv|\bm{D}|$.   Minimizing the two terms, we find the second term is most negative when (\rmnum{1}) $\theta=-\pi/2$ and (\rmnum{2}) $k=k_z$.  Condition (\rmnum{2}) also minimizes the positive-definite first term.   The remaining terms become
\begin{equation}
\label{equ: continuous case minimized Lorentz force term}
(kA+\alpha_n D)^2+(k^2-\alpha_n^2)2D^2.
\end{equation}
The marginally stable mode can thus be $\alpha_n=\pm k$ and $A=\mp D$.  This is a circularly polarized wave for the electric field.

There is in fact one more possible marginally stable eigen-mode.  That is, $\bm{k}\times\bm{A}=0$ and $\bm{k}\times\bm{D}=0$, meaning $\bm{k}$, $\bm{A}$ and $\bm{D}$ are all parallel, and therefore all three terms in Eq. (\ref{equ: continuous case Lorentz force term}) separately vanish. It corresponds to the longitudinal electric field eigen-mode.

To address the compressional contribution to the remaining energy, we first consider the longitudinal electric field.   We first sum up the Fourier components of $\bm{E}_k$. Let $\bm{E}_k = i\bm{k} g_k$, and the real-space electric field becomes
\begin{equation}
\label{equ: electric field in real space}
\bm{E}(\bm{x})= \nabla g(\bm{x}),
\end{equation}
where $g_k$ is the Fourier component of $g(\bm{x})=|\bm{B}_0|^2 h(\bf{x})$. As the electric field is only determined by the perpendicular component of the displacement, $\bm{\xi}_\perp(\bm{x})$, we can thus relate $\bm{\xi}_\perp(\bm{x})$ to $|\bm{B}_0|^2$ and $h(\bm{x})$ via
\begin{equation}
\label{equ: perpendicular displacement formula}
\bm{\xi}_\perp = \bm{B}_0\times\nabla  h + 2 h {\bm B}_0\times \nabla \ln |\bm{B}_0|.
\end{equation}

Now, $\nabla\cdot\bm{\xi} =\nabla\cdot(\bm{\xi}_\perp + \bm{\xi}_\parallel)$ and the first term becomes
\begin{equation}
\label{equ: divergence of perpendicular displacement}
\nabla\cdot\bm{\xi}_\perp = \alpha_n (\bm{B}_0\cdot \nabla h + h \bm{B}_0\cdot\nabla \ln |\bm{B}_0|^2) + (\bm{B}_0\times\nabla\ln |\bm{B}_0|^2)\cdot \nabla h.
\end{equation}
On the other hand, the second term, as before, is $\nabla\cdot \bm{\xi}_\parallel = \bm{B}_0\cdot\nabla f$.    To have zero compression, we must demand
\begin{equation}
\label{equ: divergence-free of displacement}
\bm{B}_0\cdot\nabla f +\alpha_n (\bm{B}_0\cdot \nabla h + h \bm{B}_0\cdot\nabla \ln |\bm{B}_0|^2) + (\bm{B}_0\times\nabla\ln |\bm{B}_0|^2)\cdot \nabla h=0.
\end{equation}
Having two degrees of freedom, $f$ and $h$, the solution is not unique and a solution can be obtain quite easily.  For example, let $h= \beta |\bm{B}_0|^2$, where $\beta$ is a constant, and we find
\begin{equation}
\label{equ: marginal displacement}
\bm{B}_0\cdot\nabla f = - \bm{B}_0\cdot\nabla (2\alpha_n\beta |\bm{B}_0|^2 ),
\end{equation}
yielding a regular solution $f = -2\alpha_n \beta |\bm{B}_0|^2$.

On the other hand, the minimum magnetic energy circular polarization electric field modes must have wavenumbers $k = \pm\alpha_n$, thus long-wavelength modes with wavenumbers on a spherical shell.  These modes are likely unable to make displacements compression-free and become marginally stable, and we will leave it in a future work.

In sum, the marginally stable non-trivial eigen-modes are the longitudinal electric field and trivial eigen-mode the field-aligned displacements. On the other hand, the condition for instability demands $k^2 < \alpha_n^2$, as it makes the magnetic energy negative by choosing a trial function, for example, $A=-(\alpha_n/k) D$ with $D\neq 0$.

The above provides a general result. Next we shall consider a restricted simulation space where the system is confined in a periodic cubic box, $[0,1]\times[0,1]\times[0,1]$, and the space is discrete.

Due to the periodic boundary condition, the quantity $\bm{E}$ can be expanded as $\sum_{j,l,m \in \mathbb{Z}} \bm{\tilde{E}}_{j,l,m}\exp[i 2\pi(jx+ly+mz)]$, where $\mathbb{Z}$ is the set of all integer numbers. It follows,
\begin{equation}
\label{equ: E energy term}
\left.\begin{aligned}
& \int_{\Omega} |\nabla \times (\bm{\xi}\times \bm{B}_0)|^2 d^3\bm{x} - \alpha_n \int_{\Omega} (\bm{\xi}^{*}\times\bm{B}_0)\cdot[\nabla \times (\bm{\xi}\times \bm{B}_0)] d^3 \bm{x}  \\
& = 4\pi^2 \sum_{j,l,m \in \mathbb{Z}} [|\bm{k}_{j,l,m} \times \bm{\tilde{E}}_{j,l,m}|^2 - i(\bm{k}_{j,l,m} \times \bm{\tilde{E}}_{j,l,m})\cdot \bm{\tilde{E}}_{j,l,m}^{*}]\equiv 4\pi^2 \sum_{j,l,m \in \mathbb{Z}} A_{j,l,m},
\end{aligned}
\right.
\end{equation}
where $A_{j,l,m} \equiv |\bm{k}_{j,l,m} \times \bm{\tilde{E}}_{j,l,m}|^2 - i(\bm{k}_{j,l,m} \times \bm{\tilde{E}}_{j,l,m})\cdot \bm{\tilde{E}}_{j,l,m}^{*}$ and $k_{j,l,m} \equiv (j,l,m)$.  Hereafter we will show $A_{j,l,m} \geq 0$ for any $(j,l,m)$ so the ground state ($\alpha_n=2\pi$) is stable.

Let $\bm{\tilde{E}}_{j,l,m}$ be $(e_1, e_2, e_3)$, where $e_1$, $e_2$ and $e_3$ are complex and functions of $(j,l,m)$. It follows
\begin{equation}
\label{equ: A_jlm}
A_{j,l,m} =  |le_3 - me_2|^2 + |je_3-me_1|^2 + |je_2-le_1|^2 - i[j(e_2e_3^*-e_2^*e_3) + l(e_3e_1^*-e_3^*e_1) + m(e_1e_2^*-e_1^*e_2)].
\end{equation}

We note that $A_{j,l,m}$ is $U(1)$ symmetric, i.e., $A_{j,l,m}$ remains the same under the transformation: $e_1 \rightarrow e_1\exp(i\theta)$, $e_2 \rightarrow e_2\exp(i\theta)$ and $e_3 \rightarrow e_3\exp(i\theta)$ with $\theta \in [0,2\pi)$. This property will be used in the following argument.

We decompose $(j,l,m)$ into following four cases.

\bigskip

(\rmnum{1}) $j=l=m=0$

\medskip

The quantity $A_{j,l,m}=0$ can minimize the magnetic energy and this trivial eigen-function corresponds to the field-aligned displacement.

\bigskip

(\rmnum{2}) Only one of $j$, $l$ and $m$ is non-zero

\medskip

Without loss of generality, we assume $j\neq 0$ and $l=m=0$. Therefore $A_{j,l,m}$ becomes
\begin{equation}
\label{equ: A_jlm case 2}
A_{j,l,m} = j^2|e_2|^2 + j^2|e_3|^2 - i[j(e_2e_3^*-e_2^*e_3)].
\end{equation}
By the $U(1)$ symmetry, we can let $e_2=|e_2|$ and $e_3=|e_3|exp(i\theta_3)$ with $\theta_3 \in [0,2\pi)$. Hence Eq. (\ref{equ: A_jlm case 2}) becomes
\begin{equation}
\label{equ: A_jlm case 2 estimated}
\left.\begin{aligned}
A_{j,l,m} & = j^2|e_2|^2 + j^2|e_3|^2 - 2j|e_2||e_3|\sin(\theta_3) \\
          & = j^2 \Big [ |e_2|^2 + |e_2|^2 -{2\over j}|e_2||e_3|\sin(\theta_3) \Big ] \\
          & \geq j^2\Big ( |e_2|^2 + |e_2|^2 - {2\over |j|}|e_2||e_3| \Big ) \\
          & \geq j^2(|e_2|^2 + |e_2|^2 - 2|e_2||e_3|) \\
          & = j^2 (|e_2|-|e_3|)^2 \\
          & \geq 0.
\end{aligned}
\right.
\end{equation}
Here the first inequality is based on $-1 \leq \sin(\theta_3) \leq 1$ and the second one is from $|j| \geq 1$. Therefore $A_{j,l,m}$ is always greater than zero and hence the modes are stable.

The minimum magnetic energy modes can be attained in the following two situations. First, it is achieved by $e_2=e_3=0$ and $e_1 \neq 0$. This setting makes Eq. (\ref{equ: A_jlm case 2 estimated}) be zero automatically.  Since the only non-vanishing $e_1$ is parallel to the wave propagation, it corresponds to a longitudinal electric field.

The second situation is followed by Eq. (\ref{equ: A_jlm case 2 estimated}). The equality in Eq. (\ref{equ: A_jlm case 2 estimated}) holds if and only if $|e_2|=|e_3|\neq 0$ and $j=1$ for $\theta_3=\pi/2$ or $j=-1$ for $\theta_3=-\pi/2$. This situation only permits the longest wavelength mode, i.e., $|j|=1$.  Due to the fact $|\theta_3|=\pi/2$, the condition leads to right-hand helicity, circularly polarized waves ($j=1$ for $\theta_3=\pi/2$ or $j=-1$ for $\theta_3=-\pi/2$). The minimum magnetic energy mode favors right-hand helicity because the magnetic helicity of $\bm{B}_0$ is right-handed.  If we trivially change the parameter $\alpha_n \rightarrow -\alpha_n$, everything will be left-handed.

\bigskip

(\rmnum{3}) One of $j$, $l$ and $m$ is zero.

\medskip

Again, we can consider only the case of $j=0$, $l \neq 0$ and $m \neq 0$.  As before, we now choose the phase angle such that $e_1=|e_1|$. In order to simplify the notation, we rescale $e_2$ by the factor $m$ and $e_3$ by the factor $l$, i.e. $e_2 \rightarrow me_2$ and $e_3 \rightarrow le_3$. It follows
\begin{equation}
\label{equ: A_jlm case 3}
\left.\begin{aligned}
A_{j,l,m} & = |e_2-e_3|^2 + m^2|e_1|^2 + l^2|e_1|^2 - i(e_3e_1^*-e_3^*e_1 + e_1e_2^*-e_1^*e_2 ) \\
          & = [\Re(e_2) - \Re(e_3)]^2 + [\Im(e_2) - \Im(e_3)]^2 + (m^2+l^2)|e_1|^2 + 2\Im(e_3)|e_1| -2\Im(e_2)|e_1| \\
          & = [\Re(e_2) - \Re(e_3)]^2 + [\Im(e_2) - \Im(e_3)-|e_1|]^2 + (m^2+l^2-1)|e_1|^2 \\
          & \geq 0,
\end{aligned}
\right.
\end{equation}
where $\Re$ stands for the real part and $\Im$ the imaginary part. Here the last inequality arises from $l,m \geq 1$.

The minimum magnetic energy mode is achieved by letting $e_1=0$ and $me_2=le_3$. This makes $\bm{\tilde{E}}_{j,l,m}$ parallel to the propagation direction ($(e_2, e_3) = c(l,m)$ for some constant $c$), a longitudinal electric field mode. On the other hand, the circular polarization eigen-mode in this case is not the minimum magnetic energy mode since $l$ and $m$ are non-zero and it is not the longest wavelength mode.

\bigskip

(\rmnum{4}) All of $j$, $l$ and $m$ are non-zeros

\medskip

In this case, we follow the convention in case (\rmnum{3}), i.e., $e_1=|e_1|$, $e_2 \rightarrow me_2$ and $e_3 \rightarrow le_3$. Therefore $A_{j,l,m}$ becomes
\begin{equation}
\label{equ: A_jlm case 4}
A_{j,l,m} = X^{T}MX + 2|e_1|Q^{T}X + (m^2+l^2)|e_1|^2,
\end{equation}
where
\begin{equation}
\label{equ: definition of X and Q}
X=
\begin{bmatrix}
\Re(e_2) \\
\Im(e_2) \\
\Re(e_3) \\
\Im(e_3)
\end{bmatrix}
, \text{  }
Q=
\begin{bmatrix}
-{{jl} \over m} \\
-1              \\
-{{jm} \over l} \\
1
\end{bmatrix}
,
\end{equation}
and
\begin{equation}
\label{equ: definition of M}
M=
\begin{bmatrix}
1 + \Big ( {j\over m} \Big )^2 & 0                               & -1                             & -{j\over {ml}}                 \\
0                              & 1 + \Big ( {j\over m} \Big )^2  & {j\over {ml}}                  & -1                             \\
-1                             & {j\over {ml}}                   & 1 + \Big ( {j\over l} \Big )^2 & 0                              \\
 -{j\over {ml}}                & -1                              & 0                              & 1 + \Big ( {j\over l} \Big )^2
\end{bmatrix}
.
\end{equation}
Here the superscript $T$ is the transpose of a given matrix. Since the matrix $M$ is symmetric, it can be diagonalized, i.e., $M=S\Lambda S^{T}$ with an unitary matrix $S$ and a diagonal matrix $\Lambda$. Next we let $\tilde{X}$ be $S^{T}X$ and $\tilde{Q}$ be $S^{T}Q$, Equation (\ref{equ: A_jlm case 4}) becomes
\begin{equation}
\label{equ: A_jlm case 4 next step}
A_{j,l,m} = (\tilde{X} + |e_1|\Lambda^{-1}\tilde{Q})^{T}\Lambda(\tilde{X} + |e_1|\Lambda^{-1}\tilde{Q}) + (m^2+l^2 - Q^{T}M^{-1}Q)|e_1|^2.
\end{equation}
Here the matrix $\Lambda$ has the following form
\begin{equation}
\label{equ: definition of Lambda}
\Lambda=
\begin{bmatrix}
\lambda_{+} & 0           & 0           & 0           \\
0           & \lambda_{+} & 0           & 0           \\
0           & 0           & \lambda_{-} & 0           \\
0           & 0           & 0           & \lambda_{-}
\end{bmatrix}
,
\end{equation}
with
\begin{equation}
\label{equ: definition of eigenvalue}
\lambda_{\pm} = {1\over2} \Big [ 2 + \Big ( {j \over m}\Big )^2 + \Big ( {j \over l}\Big )^2 \pm \sqrt{4 + \Big ( {{j^2}\over{m^2}} - {{j^2}\over{l^2}}\Big )^2 + \Big ( {{2j}\over{ml}}\Big )^2}\Big ].
\end{equation}
A straightforward calculation shows $Q^{T}M^{-1}Q= m^2+l^2$, meaning that the second term in Eq. (\ref{equ: A_jlm case 4 next step}) is zero. Hence Eq. (\ref{equ: A_jlm case 4 next step}) becomes
\begin{equation}
\label{equ: A_jlm case 4 next step 2}
A_{j,l,m} = (\tilde{X} + |e_1|\Lambda^{-1}\tilde{Q})^{T}\Lambda(\tilde{X} + |e_1|\Lambda^{-1}\tilde{Q}).
\end{equation}
It is noted that $\lambda_{+} \geq \lambda_{-} \geq 0$.  The positive $\Lambda$ makes the right hand side of Eq. (\ref{equ: A_jlm case 4 next step 2}) positive definite. Therefore, the quantity $A_{j,l,m}$ is always positive.

The minimal magnetic energy state can be obtained when $\tilde{X} + |e_1|\Lambda^{-1}\tilde{Q}=0$. It is equivalent to $X=-M^{-1}Q|e_1|$. A straightforward calculation yields that $\Im(e_2)=\Im(e_3)=0$, $\Re(e_2) = (l/n)|e_1|$ and $\Re(e_3) = (m/n)|e_1|$. This solution is nothing but the longitudinal electric field mode.  On the other hand, the circular-polarized mode is not the minimum magnetic energy mode, as the argument applied to case (\rmnum{3}) also applies here.

\bigskip

\medskip

Combining all above cases, it follows that Eq. (\ref{equ: E energy term}) is positive definite so the ground state ($\alpha_n=2\pi$) is stable.  The marginally stable modes that we have so far proved and that make all terms in the energy integral vanishes are the trivial mode where the displacement is parallel to the equilibrium magnetic field and the non-trivial longitudinal electric field.   The circular polarization modes are likely stable, but we so far have no proof.  Both marginally stable modes can be of small scale.  We suspect that the ideal MHD marginally stable mode may be most vulnerable to fast resistivity instability \citep*{Rosenbluth1973} in the presence of small numerical dissipation.

\label{lastpage}

\end{document}